\def\Nc{N_{\rm c}}

\def\k{{\bm k}}
\def\x{{\bm x}}

\def\Im{\operatorname{Im}}

\def\zh{z_{\rm h}}
\def\rh{r_{\rm h}}
\def\Rads{L}

\def\rstar{r_*}
\def\rstaro{r_{*0}}
\def\m{\bar m}
\def\gammat{\gamma^0}
\def\gammaz{\gamma^5}
\def\psit{\psi_0}
\def\psiz{\psi_5}

\documentclass[prd,preprint,eqsecnum,nofootinbib,amsmath,amssymb,
               tightenlines,dvips,floatfix]{revtex4}
\usepackage{graphicx}
\usepackage{bm}%	bold math
\usepackage{amsfonts}
\usepackage{amssymb}
\usepackage{dsfont}
\usepackage{undertilde}

\begin {document}

%%%%%%%%%%%%%%%%%%%%%%%%%%%%%%%%%%%%%%%%%%%%%%%%%%%%%%%%%%%%%%%%%%%%%%%%%%%%%%%
%%%%%%%%%%%%%%%%%%%%%%%%%%%%%%%%%%%%%%%%%%%%%%%%%%%%%%%%%%%%%%%%%%%%%%%%%%%%%%%

\title
    {
      Gravitino and other spin-\boldmath$\tfrac32$
      quasinormal modes in
      Schwarzschild-AdS spacetime
    }

\author{
  Peter Arnold, Phillip Szepietowski, and Diana Vaman
}
\affiliation
    {%
    Department of Physics,
    University of Virginia, Box 400714,
    Charlottesville, Virginia 22904, USA
    }%

\date {\today}

\begin {abstract}%
{%
   We investigate quasinormal mode frequencies $\omega_n$
of gravitinos and generic
massive spin-$\tfrac32$ fields in a Schwarzschild-AdS$_D$ background in
spacetime dimension $D>3$, in the black brane (large black hole)
limit appropriate
to many applications of the AdS/CFT correspondence.
First, we find asymptotic formulas for $\omega_n$ in the limit of
large overtone number $n$.  Asymptotically,
$\omega_n \simeq n \, \Delta\omega + O(\ln n) + O(n^0)$, where
$\Delta\omega$ is a known constant, and here we
compute the $O(\ln n)$ and $O(n^0)$ corrections to the
leading $O(n)$ behavior.  Then we compare to numerical calculations
of exact quasinormal mode frequencies.
Along the way, we also improve the reach and accuracy of an
earlier, similar analysis of spin-$\tfrac12$ fields.
}%
\end {abstract}

\maketitle
\thispagestyle {empty}

%%%%%%%%%%%%%%%%%%%%%%%%%%%%%%%%%%%%%%%%%%%%%%%%%%%%%%%%%%%%%%%%%%%%%%%%%%%%%%%

\section {Introduction and Results}
\label{sec:intro}

\subsection {Overview}

The original application of gauge-gravity duality was to the
conjectured equivalence of strongly-coupled large-$\Nc$
${\cal N}{=}4$ super Yang Mills gauge
theory with Type IIB supergravity in AdS$_5\times S^5$.
For studies of the gauge theory at finite temperature, a black
brane is introduced in the gravity theory, replacing
AdS$_5$ by Schwarzschild-AdS$_5$ (hereafter SAdS$_5$).
This theory is often used as a model to study strongly-coupled
QCD-like quark-gluon plasmas.
In gauge-gravity duality,
corrections of order $1/\Nc^2$ in the large-$\Nc$ field
theory correspond to loop corrections in the gravity theory,
and there is a beautiful and very general expression by
Denef, Hartnoll, and Sachdev \cite{DHS} that relate these
to quasinormal mode frequencies of all the various fields in the
gravity theory.%
\footnote{
   For a discussion of larger $O(\lambda^{1/2}/\Nc^2)$ corrections,
   where $\lambda$ is the 't Hooft coupling constant, see
   refs.\ \cite{GKT,MPS}.
}
Their formula has been applied to a
variety of examples in gauge-gravity duality but has yet to
be fully implemented in the case of SAdS$_5$.
Though there has been a great deal of numerical and analytic work
over the years on quasinormal modes of black holes,%
\footnote{
  For some modern reviews, see refs.\ \cite{QNMreview1,QNMreview2}.
}
there are a few relevant cases yet to be filled in.
In particular, there has not been a complete analysis of
quasinormal modes in SAdS$_5$ of all the different types of
fields that arise in the compactification of type IIB supergravity
from SAdS$_5 \times S^5$.  It is our intention to fill in those
missing cases.  In this particular paper, we analyze the quasinormal mode
frequencies $\omega_n$ of the gravitino, and
more generally of spin-$\tfrac32$ fields of any mass, in SAdS$_D$
for spacetime dimension $D > 3$.
We will find analytic asymptotic formulas for
large overtone number $n$ (i.e.\ large $|\omega_n|$).  For
the specific case $D{=}5$, we will compare these results
to accurate numerical calculations of exact quasinormal mode
frequencies.
The analysis will generalize recent work on
the spin-$\tfrac12$ case \cite{dirac} in SAdS$_D$, which we will also improve
upon.  For large enough $n$, the asymptotic formula will have the
schematic form
\begin {equation}
   \omega_n \simeq n \, \Delta\omega + A \ln n + B + \cdots ,
\label {eq:schematic}
\end {equation}
where $\Delta\omega$ is the asymptotically constant spacing between
successive modes, and $B$ depends logarithmically on the
spatial momentum $\k$ parallel to the boundary.
We will find $\omega_n$ through $O(n^0)$.  We will also find a more
general asymptotic formula whose validity extends to
lower values of $n$ than does the schematic form (\ref{eq:schematic}).

In the remainder of this introduction, we first very
briefly review the form of the Dirac and Rarita-Schwinger equations
for spin-$\tfrac12$ and $\tfrac32$ fields in SAdS.
Then we preview our results, presenting analytic formulas for
asymptotic quasinormal mode
frequencies together with a comparison to numerical results for exact
quasinormal mode frequencies (see fig.\ \ref{fig:wplane}).
In section \ref{sec:dirac}, we then review and rederive earlier
asymptotic results \cite{dirac} for the spin-$\tfrac12$ case.
This rederivation serves two purposes.  First, some of the details
of the derivation are different from ref.\ \cite{dirac}
and provide a simpler starting point for our generalization to the
spin-$\tfrac32$ case.  Second, the new derivation improves on the
previous result, extending the range of validity from asymptotically
large values of $n$ to more moderate values of $n$, and also to
the case of vanishing spatial momentum $\k$ (which was not covered
by the result of ref.\ \cite{dirac}).
With these preliminaries out of the way,
section \ref{sec:spin32} generalizes the method in order to find
asymptotic results for the spin-$\tfrac32$ case.
Finally, in section \ref{sec:numerics}, we present our technique for
numerical calculation of exact quasinormal mode frequencies in the
spin-$\tfrac32$ case, and we present a more accurate test of
asymptotic formulas vs.\ numerics.

% -------------------------------------------------------------------------

\subsection{Metric and Field Equations}

We will use the form
\begin {equation}
   ds^2 = \frac{\Rads^2}{z^2} \left[ -f \, dt^2 + d\x^2 + f^{-1} dz^2 \right]
\label {eq:metric}
\end {equation}
for the SAdS$_D$ in $D$ spacetime dimensions in the black brane
(large black hole) limit.
Here, $\Rads$ is the radius associated with the asymptotic AdS spacetime,
and%
\footnote{
   Many papers in the literature on quasi-normal modes use the letter $f$ to
   instead refer to $r^2$ times our $f$, where $r = \Rads^2/z$.
}
\begin {equation}
   f
   = 1 - \Bigl( \frac{z}{\zh} \Bigr)^{D-1} .
\end {equation}
The boundary of AdS is at $z=0$, the black hole horizon
is at $z=\zh$, and the singularity is at $z=\infty$.
We will refer to the spacetime dimension of the
boundary of SAdS$_D$ as $d = D-1$.
The relationship between $\zh$ and the Hawking temperature $T$ is
\begin {equation}
   T = \frac{D-1}{4\pi \zh} = \frac{d}{4\pi \zh} \,.
\label {eq:T}
\end {equation}
For derivations in this paper, we will often adopt units where
$\zh=1$.

We will consider both spin-$\tfrac12$ and spin-$\tfrac32$
fermions of mass $m$ propagating in this metric background.
For spin-$\tfrac12$ fermions, the equation of motion is the
curved-space Dirac equation
\begin {equation}
   \Gamma^M D_M \Psi - m \Psi = 0,
\label{eq:diraceq}
\end {equation}
where $\Gamma^M$ are curved space Dirac matrices, with
$\{\Gamma^M,\Gamma^N\} = 2g^{MN}$,
and where the covariant derivative $D_M$ contains a
spin connection term.
For spin-$\tfrac32$ fermions, the equation is%
\footnote{
  Some authors use a different sign convention for the mass terms in the
  equations of motion.  Different sign conventions are physically
  equivalent since
  the replacement $\Gamma^M \to -\Gamma^M$ generates an equally valid
  representation of the $\Gamma$ matrices.
}
\cite{GravitinoEq}
\begin {equation}
   \Gamma^{MNP} D_N \Psi_P + m \Gamma^{MN} \Psi_N = 0 ,
\label {eq:spin32eq}
\end {equation}
where $\Gamma^{MNP} \equiv \Gamma^{[M} \Gamma^N \Gamma^{P]}$ and
$\Gamma^{MN} \equiv \Gamma^{[M} \Gamma^{N]}
  = \tfrac12(\Gamma^M \Gamma^N {-} \Gamma^N \Gamma^M)$
are anti-symmetrized
products of $\Gamma$ matrices.%
\footnote{
  In \cite{GravitinoEq} the massive spin 3/2 equation (\ref{eq:spin32eq})
  is written using the epsilon symbol in a form specific to 4 spacetime
  dimensions, but the $D$-dimensional equation quoted in (\ref{eq:spin32eq})
  can be easily inferred.  A more general equation, allowing for a mass term
  of the type $m' \Psi_M$ to be added to (\ref{eq:spin32eq}) was considered
  in \cite{corley,kosh,vis}. This was in part motivated by considering a
  Kaluza-Klein reduction of IIB supergravity on $S^5$. However, as shown by
  Kim, Romans and van Nieuwenhuizen \cite{KRvN}, in the lower dimensional
  theory the spin 3/2 fields are either the gravitino, with a specific
  non-zero mass (a consequence of a non-zero cosmological constant resulting
  from the reduction on $S^5$) or massive fields constrained to obey
  $D^M \varphi_M=0$ and with an equation of motion of the type
  $\Gamma^{MNP} D_N \varphi_P + m' \varphi^M=0$. In the latter case one
  can show that the equation of motion and constraint are equivalent to
  those derived starting from (\ref{eq:spin32eq}). Therefore if either
  $m$ or $m'$ are zero, one ends up solving the same type of equation
  of motion and constraints.
}

A special case of interest will be the gravitino.
In (S)AdS$_D$, it has a non-zero mass $m$ given by
\cite {GravitinoEq, GravitinoEqMassGauge}%
\footnote{
  For results valid in a generic $D$-dimensional spacetime, see for
  example ref.\ \cite{GravitinoEqMassGauge1}. However, we caution
  the reader that their conventions used in defining the spin connection
  and curvature are not the same as ours.
}
\begin {equation}
   (m\Rads)_{\rm gravitino} = \frac{D-2}{2} = \frac{d-1}{2} \,.
\label {eq:mgravitino}
\end {equation}
For this mass, the spin-$\frac32$ equation (\ref{eq:spin32eq})
in (S)AdS has a gauge symmetry \cite{GravitinoEq, GravitinoEqMassGauge},
\begin {equation}
   \delta\Psi_M = \left(D_M - \tfrac1{2\Rads}\, \Gamma_M\right)\epsilon  ,
\label {eq:gauge}
\end {equation}
where $\epsilon(x)$ is an arbitrary Dirac-spinor-valued function.
We will study the gravitino in $\Gamma^M \Psi_M{=}0$ gauge.
It will be useful for later to note that in this gauge, the
equation of motion for the gravitino ghost is the Dirac
equation (\ref{eq:diraceq}) with mass%
\footnote{
  The ghost equation of motion is determined by
  the gauge transformation of the gauge constraint function:
  $\delta(\Gamma^M \Psi_M) = \Gamma^M \delta\Psi_M =
   (\Gamma^M D_M - \tfrac{d+1}{2\Rads})\epsilon$.
  See Appendix \ref{app:ghost} for a more thorough discussion of the
  gauge and ghost degrees of freedom.
}
\begin {equation}
   (m\Rads)_{\rm ghost} = \frac{D}{2} = \frac{d+1}{2} \,.
\end {equation}

In applications of gauge-gravity duality,
masses $m$ of the spin-$\tfrac12$ and spin-$\tfrac32$
fields in the gravity theory
are related by duality
to conformal dimensions $\Delta$ of spin-$\tfrac12$ and
spin-$\tfrac32$ operators in the field
theory by $|m\Rads| = \Delta - \frac{d}{2}$ \cite{MAGOO}.
%% section 3.3.1

Throughout, we will adopt the convention that the mass $m$
is non-negative.  We will also focus on the case of non-zero $m$.
In the case of Type IIB supersymmetry on (S)AdS$_5\times S^5$, for
example, all of the fermion fields have non-zero mass after compactification
on the $S^5$ \cite{KRvN}.

% -------------------------------------------------------------------------

\subsection{Results}

\subsubsection{Improved spin-$\tfrac12$ results}

For comparison, we first give the result for
(retarded) quasinormal mode frequencies for spin-$\frac12$ particles
in $D>3$,
which will be reviewed in section \ref{sec:dirac}:
\begin {equation}
   \frac{\omega_{n}}{\pi T} \simeq
   4 e^{-i \pi/d} \sin(\tfrac{\pi}{d})
   \left[
     n
     + \frac{m\Rads}{2}
     + \frac{i}{2\pi} \ln \left(
          i \lambda \, \frac{k/\pi T}{(\omega/\pi T)^a}
          - \Lambda \, \frac{m\Rads}{(\omega/\pi T)^{1-a}}
       \right)
  \right]_{\omega/\pi T = 4 e^{-i\pi/d}\sin(\frac{\pi}{d})n} ,
\label {eq:dirac}
\end {equation}
where
\begin {subequations}
\label {eq:lLambda}
\begin {equation}
   \lambda \equiv
   e^{-i\pi a/2}(1-a)^{1-a} \, \Gamma(a) \, \sin(\pi a) ,
\end {equation}
\begin {equation}
   \Lambda \equiv
   \frac{2}{(d-1)} \, e^{-i\pi(1-a)/2} (1-a)^{-(1-a)} \, \Gamma(1-a) \, \sin(\pi a)
   = - \frac{2\pi i \sin(\pi a)}{(d-1)\lambda}
   \,,
\end {equation}
\end {subequations}
and
\begin {equation}
   a \equiv \frac{d-2}{2(d-1)} \,.
\label {eq:a}
\end {equation}
Above, $k \equiv \pm |\k|$, where $\k$ is the $(d-1)$-dimensional
spatial momentum and the sign $\pm$ is determined by the spin state of
the fermion, in a way that will be stated precisely in
section \ref{sec:dirac}.
The above formula gives retarded quasi-normal mode frequencies
in the lower-right quadrant of the complex frequency plane.
The corresponding frequencies in the lower-left quadrant are
given by $\omega = -\omega_n^*(k\to -k)$.%
\footnote{
  A few more technical notes:
  $\pm 2\pi i$ ambiguities in
  the value of the logarithm in (\ref{eq:dirac}) may
  be absorbed into a redefinition of the overtone
  number $n$.
  The quasinormal modes here assume the typical gauge-gravity duality
  boundary condition
  that the field vanishes at the boundary, which implicitly requires
  $m \not= 0$.
  For a discussion of other boundary conditions and $m=0$,
  see refs.\ \cite{dirac,GiammatteoJing}.
}

Note that the substitution indicated in
(\ref{eq:dirac}) for $\omega/\pi T$ in the argument of the logarithm
is just the leading $O(n)$ result for $\omega_n/\pi T$.
One could instead drop this substitution and solve (\ref{eq:dirac})
self-consistently for $\omega=\omega_n$, but the difference would be
$O(n^{-1})$ and so beyond the order to which we have calculated.
In any case, the substitution introduces a logarithmic dependence
on $n$.

The $k$ term in the argument of the
logarithm in (\ref{eq:dirac}) dominates over the $m$ term for
large enough $n$
(because $a < 1-a$).  Specifically, for $n \gg (m\Rads T/k)^{d-1}$,
the formula simplifies to
\begin {equation}
   \frac{\omega_{n}}{\pi T} \simeq
   4 e^{-i \pi/d} \sin(\tfrac{\pi}{d})
   \left[
     n
     + \frac{m\Rads}{2}
     + \frac{i}{2\pi} \ln \left(
          i \lambda \, \frac{k/\pi T}{(\omega/\pi T)^a}
       \right)
  \right]_{\omega/\pi T = 4 e^{-i\pi/d}\sin(\frac{\pi}{d})n} .
\end {equation}
This is equivalent to the asymptotic formula that was found in
ref.\ \cite{dirac}.  As we will discuss, the more complicated
formula (\ref{eq:dirac}) has the advantage that it is
accurate for
a wider range of $n$ when $k$ is small and in particular
handles the case $k=0$.

% .........................................................................

\subsubsection{Spin-$\tfrac32$ results}

There are three different branches of quasi-normal mode results
for the spin-$3/2$ case, relative to the spin-$1/2$ case.
As we will discuss later, there are $d-3$ transverse polarizations of
$\Psi_M$ which each reduce to the Dirac case with mass $m$
\cite{Policastro,Erdmenger}.
These then have
asymptotic quasinormal mode frequencies given by (\ref{eq:dirac}).
There are also two other polarizations, for which we find
\begin {subequations}
\label {eq:WKBbranches}
\begin {multline}
   \frac{\omega^{(1)}_{n}}{\pi T} \simeq
   4 e^{-i \pi/d} \sin(\tfrac{\pi}{d})
\times \\
   \left[
     n
     + \frac{m\Rads}{2}
     + \frac{i}{2\pi} \ln \left(
        \frac{(d-3)}{(d-1)} \, i \lambda \, \frac{k/\pi T}{(\omega/\pi T)^a}
        - \Lambda \, \frac{m\Rads}{(\omega/\pi T)^{1-a}}
      \right)
  \right]_{\omega/\pi T = 4 e^{-i\pi/d}\sin(\frac{\pi}{d})n}
\label {eq:WKBbranch1}
\end {multline}
and
\begin {multline}
   \frac{\omega^{(2)}_{n}}{\pi T} \simeq
   4 e^{-i \pi/d} \sin(\tfrac{\pi}{d})
\times \\
   \left[
     n
     + \frac{m\Rads}{2}
     + \frac{i}{2\pi} \ln \left(
        - i \lambda \, \frac{k/\pi T}{(\omega/\pi T)^a}
        + \frac{(d+1)}{(d-1)} \, \Lambda \, \frac{m\Rads}{(\omega/\pi T)^{1-a}}
      \right)
  \right]_{\omega/\pi T = 4 e^{-i\pi/d}\sin(\frac{\pi}{d})n} .
\label {eq:WKBbranch2}
\end {multline}
\end {subequations}
Note that $d=3$ ($D=4$) is a special case for the logarithm in
the first branch (\ref{eq:WKBbranch1}), as there is then no dependence on $k$.
%%The transverse branches, plus the two branches above, account for
%%$2(d-1)=2(D-2)$ physical degrees of freedom for a generic, massive
%%spin-$\frac32$ field, where the factor of 2 is for the sign of
%%$k = \pm|\k|$.

As mentioned earlier, the special case of the gravitino in
AdS$_{d+1}$ corresponds to a ``mass'' of
$m\Rads=(d-1)/2$ in the
spin-$\frac32$ equation of motion (\ref{eq:spin32eq}).
In this particular case, the second branch of
non-transverse solutions (\ref{eq:WKBbranch2})
is an unphysical artifact, and only the first branch (\ref{eq:WKBbranch1})
and the $d{-}3$ transverse
polarizations turn out to be physical.
%%The gravitino has only $2(d-2)=2(D-3)$ physical degrees of freedom.
For the gravitino, the unphysical second-branch frequencies $\omega_n^{(2)}$
are the
same as the frequencies of the gravitino ghost, which are those of a
spin-$\frac12$ particle with mass $(m\Rads)_{\rm ghost} = (d+1)/2$.
One may verify this identity from the asymptotic approximations
(\ref{eq:dirac}) and (\ref{eq:WKBbranch2}), but it is also true
of the exact values of the frequencies.
See Appendix \ref{app:ghost} for a more thorough discussion of the
cancellation between ghost and unphysical-branch
quasinormal mode contributions in the gravitino
partition function.

\begin {figure}
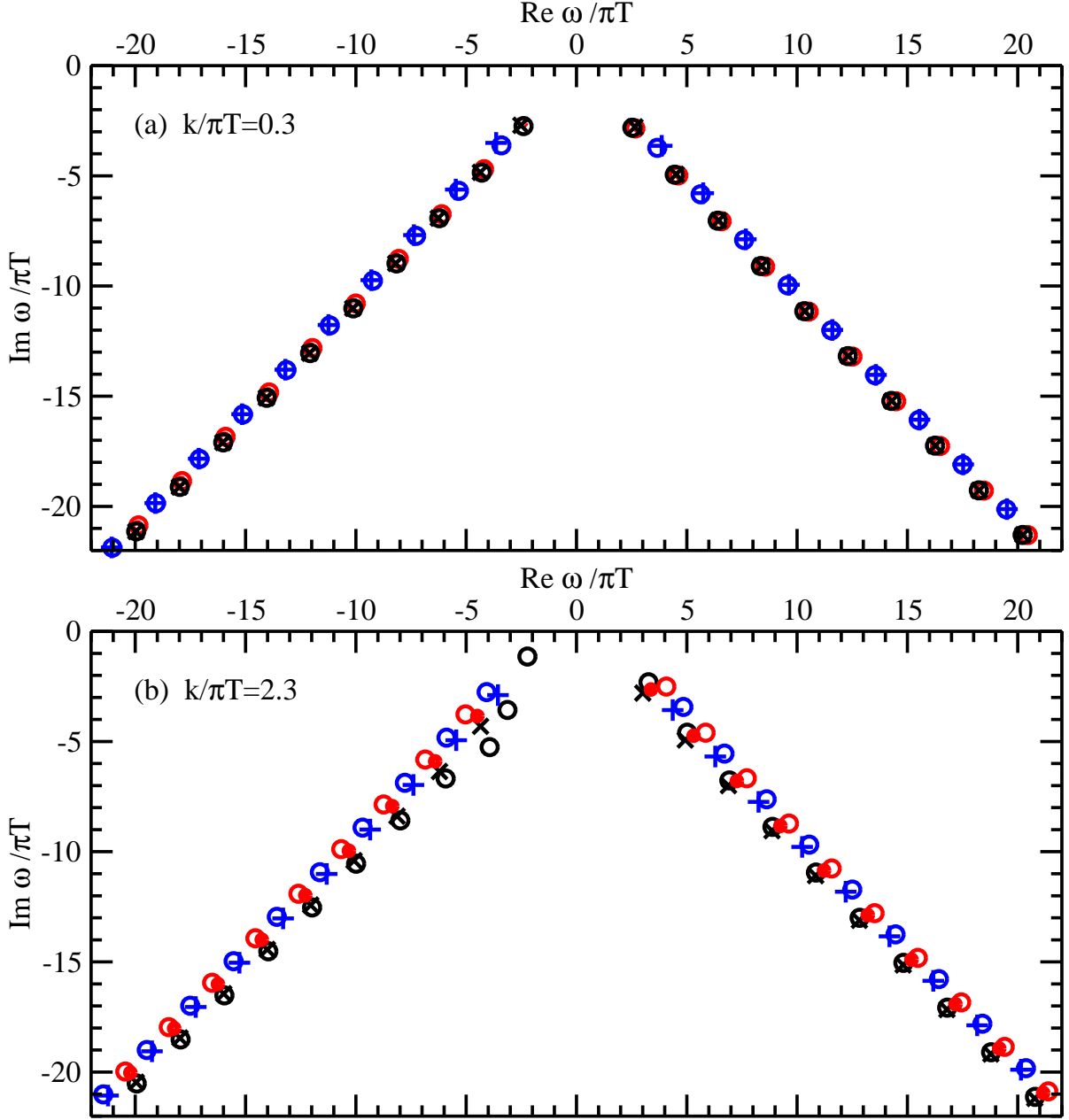

\begin {center}
  \includegraphics[scale=0.75]{wplane.eps} \\
  \includegraphics[scale=0.75]{wplane2p3.eps}
  \caption{
     \label{fig:wplane}
     Retarded spin-$\frac32$ quasi-normal mode frequencies in the
     complex $\omega$ plane for $D{=}5$, $m\Rads=\tfrac32$,
     and spin states with
     ${\bm\sigma}\cdot{\bm\hat\k} = +1$ (as defined in section
     \ref{sec:spin}).
     The two figures take (a) $|\k|=0.3\,\pi T$ and
     (b) $|\k|=2.3\,\pi T$.
     The figures are not left-right symmetric,
     and the results for ${\bm\sigma}\cdot{\bm\hat\k} = -1$ are the
     left-right mirror images of these figures, as required by parity.
     The open circles are precise numerical results,
     the small red filled circles show the asymptotic formula
     (\ref{eq:dirac}) for the transverse polarizations,
     the black $\times$'s show the
     asymptotic formula (\ref{eq:WKBbranch1}) for the first branch of
     non-transverse solutions, and
     the blue $+$'s show the asymptotic formula
     (\ref{eq:WKBbranch2}) for
     the second branch.
     The latter branch is unphysical for the particular case here
     since $m\Rads=\tfrac32$ is the gravitino case in $D{=}5$,
     and we have checked numerically
     that the corresponding blue circles exactly match
     the quasi-normal modes of the gravitino ghost---a Dirac field with
     mass $(m\Rads)_{\rm ghost} = \frac52$.
  }
\end {center}
\end {figure}

Fig.\ \ref{fig:wplane} gives a first look at the comparison of our
numerical results (described later) for exact quasi-normal mode\
frequencies and the asymptotic formulas (\ref{eq:WKBbranches})
for $D{=}5$, $m\Rads=\tfrac32$, and $k=0.3\,\pi T$ and $2.3\,\pi T$.
The asymptotic formula works well, and in the case of relatively
small $k$ works well even for low overtone number $n$.

% ..........................................................................

\subsubsection {Other cases of spin-$\tfrac32$ results}

In terms of analytic results, this paper focuses on the asymptotic
behavior for large $|\omega_n|$.  For the opposite limit of very small
$\omega$ (and $\k$), an analysis
of the supersymmetric hydrodynamic mode associated with the gravitino
may be found in refs.\ \cite{Policastro,Erdmenger}
(see also ref.\ \cite{Gauntlett}).

All of our results assume $D>3$.  The special case $D=3$ corresponds
to Banados-Teitelboim-Zanelli (BTZ) black holes \cite{BTZblackhole},
for which exact
analytic results are known for quasinormal mode frequencies.
For a discussion of results
in this case for arbitrary half-integer spin, see ref.\ \cite{BTZdatta}.

See refs.\ \cite{gravitino1,gravitino2,gravitino3} for
a variety of analytic and numerical results for
gravitino quasinormal modes of $D{=}4$
asymptotically-flat black holes.
In contrast, our work here is for asymptotically-AdS black holes,
relevant to gauge-gravity duality.

% =========================================================================

\section{Review and Improvement of Spin \boldmath$1/2$ Asymptotic Results}
\label {sec:dirac}
\label {sec:spin}

In order to address the spin-$\tfrac32$ case, it will be useful to first
review the spin-$\tfrac12$ case treated in ref.\ \cite{dirac}.
Here, we will give a treatment that is somewhat simplified compared
to ref.\ \cite{dirac} and which will be easier to generalize to
the spin-$\tfrac32$ case.%
\footnote{
   More specifically, in this paper we will work directly with the
   first-order equation of motion coming from the Dirac equation,
   whereas ref.\ \cite{dirac} worked with a second-order
   Schr\"odinger-like equation of motion derived from the Dirac equation.
}
In the process, we will also generalize earlier results to include
the $m$ dependence in the argument of the logarithm in
the asymptotic result (\ref{eq:dirac}).

As reviewed in ref.\ \cite{dirac}, the Dirac equation
(\ref{eq:diraceq}) may be rewritten as
\begin {equation}
    \left[
        \sqrt{-g^{tt}} \, \gamma^t \partial_t
        + \sqrt{g^{xx}} \, \gamma^i \partial_i
        + \sqrt{g^{zz}} \, \gamma^z \partial_z
        - m
    \right] \psi = 0 ,
\label {eq:Dirac}
\end {equation}
where the original field $\Psi$ has been rescaled
as
\begin {equation}
   \Psi \equiv (-g g^{zz})^{-1/4} \psi \propto z^{d/2} f^{-1/4} \psi
\label {eq:PsipsiDirac}
\end {equation}
and the $\gamma^m$ are flat-spacetime $\gamma$ matrices with
$\{\gamma^m,\gamma^n\} = 2 \eta^{mn}$.
From now on, we will refer to $\gamma^z$ as ``$\gammaz$,''
just because we find that notation more evocative, but our analysis will
nonetheless apply generally to arbitrary dimension $D > 3$.

Throughout this paper, our convention for labels for coordinate indices
will be that capital letters $M,N,\cdots$ refer to the $D$ curved spacetime
dimensions in SAdS$_D$; Greek letters $\mu,\nu,\cdots$ refer to the
$d=D{-}1$ spacetime dimensions parallel to the boundary; the lower-case
letters $i,j$ refer to
the $d{-}1$ spatial dimensions parallel to the boundary; and lower-case
letters $m,n,\cdots$ from roughly
the second half of the alphabet refer to the $D$ flat spacetime coordinates
of a vielbein, such as appears in the expression $\gamma^m$ above.

Multiplying (\ref{eq:Dirac}) by $-\gammaz \sqrt{-g_{tt}}$, and using the
explicit metric (\ref{eq:metric}), we will write the Dirac
equation as
\begin {subequations}
\label {eq:diracpsi}
\begin {equation}
   {\cal D}_{1/2} \psi = 0 ,
\end {equation}
where
\begin {equation}
   {\cal D}_{1/2} \equiv
      - f \partial_z
      + i \gammaz \gammat \omega
      - i f^{1/2} \gammaz {\bm\gamma}\cdot\k
      + \m \, \frac{f^{1/2}}{z} \, \gammaz
\label {eq:D1/2}
\end {equation}
\end {subequations}
and
\begin {equation}
  \m \equiv m\Rads .
\end {equation}
We will find it convenient to work in a particular basis for the
$\gamma$ matrices where
\begin {equation}
  \gammat =
    i \begin{pmatrix} 0 & \openone \\ \openone & 0 \end{pmatrix}
    = i \tau_1 \otimes \openone ,
  \qquad
  {\bm\gamma} =
    \begin{pmatrix} {\bm\sigma} & 0 \\ 0 & -{\bm\sigma} \end{pmatrix}
    = \tau_3 \otimes {\bm\sigma} ,
  \qquad
  \gammaz =
    \begin{pmatrix} 0 & -i\openone\\ i\openone & 0 \end{pmatrix}
    = \tau_2 \otimes \openone ,
\label {eq:basis}
\end {equation}
so that%
\footnote{
  For readers comparing to the earlier spin-$\tfrac12$ analysis
  in ref.\ \cite{dirac}:
  the choice of basis here is that same as the one in that paper's
  section III
  but different from the one in its section II.
  Note also that the $\eta^{(\pm)}$ of (\ref{eq:WKB1/2}) are eigenvectors
  of $\gammaz \gammat$, whereas the notation $\psi_\pm$ in
  ref.\ \cite{dirac} was used for the components corresponding to
  eigenvectors of $\gammaz$.
}
\begin {equation}
   {\cal D}_{1/2} =
      - f \partial_z
      + i \omega \tau_3
      + k f^{1/2} \tau_1
      + \m \, \frac{f^{1/2}}{z} \, \tau_2 ,
\end {equation}
where we define
\begin {equation}
   k \equiv \k\cdot{\bm\sigma} .
\end {equation}
We henceforth focus on eigenstates of $\k\cdot{\bm\sigma}$, so that
\begin {equation}
   k = \pm |\k| .
\end {equation}

We define (retarded) quasinormal mode solutions to be solutions that
simultaneously (i) vanish at the boundary ($z{=}0$) of SAdS and
(ii) are purely infalling at the horizon.

\begin {figure}
\begin {center}
  \includegraphics[scale=0.75]{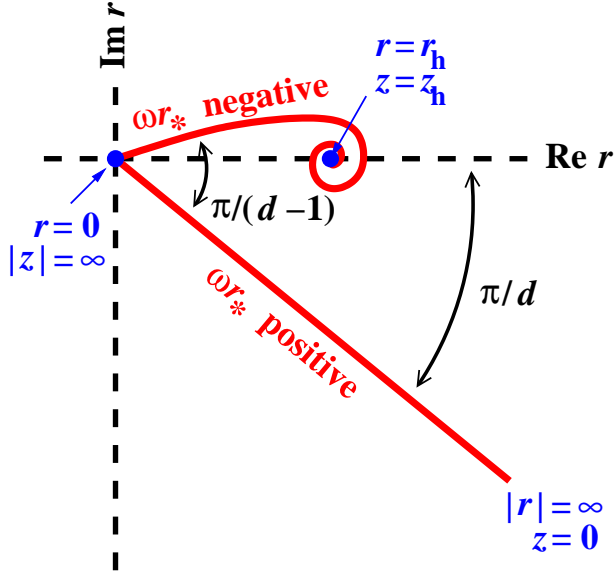}
  \caption{
     \label{fig:stokes}
     A qualitative picture of the relevant Stokes lines $\Im(\omega\rstar)=0$
     in the complex $r$ plane
     for following WKB between the boundary
     ($r=\infty$) and the horizon ($r=\rh$).
     The path passes through the singularity ($r=0$) in SAdS$_{d+1}$.
     The other Stokes lines emanating from the origin are not shown, one
     of which escapes to $-e^{-i\pi/d}\infty$ and the others which
     spiral into the complex-valued horizons $r = e^{-i2n\pi/d}\rh$
     for $n=1,\cdots,d{-}1$.
     (The spiral into the horizon shown above crosses a cut in the
     definition of $\rstar$ emanating
     from $r=\rh$, and the curve spirals onto higher and higher
     Riemann sheets.)  Given our retarded convention for $\omega$,
     this figure is the complex conjugate of similar diagrams in
     refs.\ \cite{NatarioSchiappa,CardosoNatarioSchiappa}.
  }
\end {center}
\end {figure}

To find the asymptotic quasinormal mode frequencies, we follow
ref.\ \cite{dirac} and use the
Stokes line method nicely reviewed in ref.\ \cite{NatarioSchiappa}.
Start by taking the naive large-$\omega$ limit of (\ref{eq:diracpsi}),
which is
\begin {equation}
   \left[-f \partial_z + i \omega \tau_3 \right] \psi
   \approx 0 ,
\label {eq:naive}
\end {equation}
with solution
\begin {equation}
   \psi = e^{-i \omega \rstar \tau_3} \eta
   = e^{-i\omega\rstar} \eta^{(+)} + e^{+i\omega\rstar} \eta^{(-)} .
\label {eq:WKB1/2}
\end {equation}
Here $\eta$ is a constant Dirac spinor,
$\eta^{(+)}+\eta^{(-)}$ is its decomposition into spinors $\eta^{(\pm)}$
with definite values $\pm$ of $\gammaz \gammat = \tau_3$, and
$\rstar$ is the tortoise coordinate defined by
\begin {equation}
   \rstar = \int_z^\infty \frac{dz}{f} \,.
\label {eq:rstar}
\end {equation}
As shown, the solution (\ref{eq:WKB1/2}) has components that behave as
$\exp(\mp i\omega\rstar)$.  In order to avoid having one of these component
become exponentially small, and so get lost in the approximation error of the
other component, we perform WKB by following
Stokes lines, defined by
$\Im(\omega\rstar)=0$  in the complex $z$ plane.
For asymptotic quasinormal mode frequencies, the relevant Stokes lines
are depicted qualitatively in fig.\ \ref{fig:stokes} in the
complex $r$ plane, where $r \equiv \Rads^2/z$.
To connect the quasinormal mode condition at the boundary
of SAdS with the condition at the horizon, we follow the Stokes lines from
the boundary at $z{=}0$ to the singularity at $z{=}\infty$ and
from there to the horizon at $z{=}\zh$.
The WKB solution (\ref{eq:WKB1/2}) is not valid
very close to the boundary or to the singularity
(where the $m$ and/or $k$ terms in the Dirac equation become
important even for large $\omega$), and so one has
to separately solve the Dirac equation in those limiting cases in order
to match to the WKB solutions.
Far away from the boundary and the singularity, we will refer to the
large-$\omega$ WKB solutions (\ref{eq:WKB1/2}) along the positive and
negative $\omega\rstar$ Stokes lines as
\begin {subequations}
\label {eq:WKBposneg}
\begin {equation}
   \psi = e^{-i \omega \rstar \tau_3} \eta_{\omega\rstar>0}
   = e^{-i \omega \rstar} \eta^{(+)}_{\omega\rstar>0}
     + e^{+i \omega \rstar} \eta^{(-)}_{\omega\rstar>0}
\label {eq:WKBpos}
\end {equation}
and
\begin {equation}
   \psi = e^{-i \omega \rstar \tau_3} \eta_{\omega\rstar<0}
   = e^{-i \omega \rstar} \eta^{(+)}_{\omega\rstar<0}
     + e^{+i \omega \rstar} \eta^{(-)}_{\omega\rstar<0} ,
\label {eq:WKBneg}
\end {equation}
\end {subequations}
respectively.

Along the negative $\omega\rstar$ Stokes line of
fig.\ \ref{fig:stokes}, the WKB solution (\ref{eq:WKBneg}) should have
only an $e^{-i\omega\rstar}$ component, so that it will be purely
infalling at the horizon.
So $\eta^{(-)}_{\omega\rstar<0}$ must vanish in (\ref{eq:WKBneg}).
Along the positive $\omega\rstar$ Stokes
line of the figure, the WKB solution will need both
$e^{-i\omega\rstar}$ and $e^{+i\omega\rstar}$ components in order for
$\Psi$ to be able to satisfy the other quasi-normal mode condition
that it vanish at the boundary $z{=}0$ of SAdS.
The only way both conditions can be satisfied is for the matching
near the singularity to mix $e^{-i\omega\rstar}$ and $e^{+i\omega\rstar}$
components as one moves from the positive $\omega\rstar$ Stokes lines
to the negative one.  Since the naive solution (\ref{eq:WKB1/2}) is
exact if $k=m=0$, such mixing can only occur if we study the effect
of $k$ and $m$ on the solution near the singularity.

In order to study solutions more general than the naive approximation
(\ref{eq:WKB1/2}), it will be convenient to
elevate the $\eta$ in this solution to a function of $z$.
Plugging (\ref{eq:WKB1/2}) into (\ref{eq:diracpsi}) then recasts the
Dirac equation as
\begin {equation}
  -f e^{-i\omega \rstar \tau_3} \partial_z \eta
  + f^{1/2} \left( k \tau_1 + \frac{\m}{z} \, \tau_2 \right)
    e^{-i\omega\rstar\tau_3} \eta
  = 0 ,
\end {equation}
or equivalently
\begin {equation}
  \partial_{\rstar} \eta =
  - f^{1/2} e^{+2i\omega\rstar\tau_3}
  \left( k \tau_1 + \frac{\m}{z} \, \tau_2 \right) \eta .
\label {eq:etaeq}
\end {equation}

% ----------------------------------------------------------------------------

\subsection{Behavior near the singularity}

Following ref.\ \cite{dirac}, we will match WKB solutions between the
positive and negative $\omega\rstar$ Stokes lines of fig.\ \ref{fig:stokes}
by working in a region that is close enough to the singularity that
we may make large-$z$ (small-$r$) approximations, but far enough
from the singularity that the $k$ and $m$ terms in the Dirac equation
cause only small perturbations to the naive large-$\omega$ solution
(\ref{eq:WKB1/2}).  (See ref.\ \cite{dirac} for a detailed justification,
which requires $D>3$.)
Working in units where $\zh=1$ and using the near-singularity
approximations
\begin {equation}
  f = 1-z^d \simeq -z^d ,
\end {equation}
\begin {equation}
  f^{1/2} \simeq -i z^{d/2} ,
\end {equation}
(\ref{eq:etaeq}) becomes
\begin {equation}
  \partial_{\rstar} \eta \simeq
  i z^{d/2} e^{+2i\omega\rstar\tau_3}
  \left( k \tau_1 + \frac{\m}{z} \, \tau_2 \right) \eta .
\end {equation}
In earlier work \cite{dirac} we dropped the $\m/z$ term compared
to the $k$ term, but now we will keep it in order to improve the
approximation.  (As we will discuss later, this improvement will extend
the validity of our asymptotic formula to a wider range of overtone
numbers $n$ when $|\k| \ll \m^{(d-2)/d} T$.)
In order to treat $k$ and $\m$ perturbatively, write
\begin {equation}
   \eta(z) = \bar\eta + \delta\eta(z) ,
\label {eq:etax}
\end {equation}
where $\bar\eta$ is constant, and expand to first order in $\delta\eta$,
$k$ and $\m$:
\begin {equation}
  \partial_{\rstar} (\delta\eta) =
  i z^{d/2} e^{+2i\omega\rstar\tau_3}
  \left( k \tau_1 + \frac{\m}{z} \, \tau_2 \right) \bar\eta .
\label {eq:deta1}
\end {equation}
Then
\begin {equation}
  \delta\eta =
  -i \int_{\rstar}^\infty d\rstar \> z^{d/2} e^{+2i\omega\rstar\tau_3}
  \left( k \tau_1 + \frac{\m}{z} \, \tau_2 \right) \bar\eta .
\label {eq:detaint}
\end {equation}
Near the singularity, $z$ is related to $\rstar$ by
\begin {equation}
  \rstar = \int_z^\infty \frac{dz}{f} \simeq - \frac{1}{(d-1) z^{d-1}} ,
\end {equation}
and so
\begin {equation}
  z \simeq \left[ e^{-i\pi}(d-1)\rstar \right]^{-1/(d-1)} .
\label {eq:zrstar}
\end {equation}
The result of the $\rstar$ integration in (\ref{eq:detaint}) is
then
\begin {equation}
  \delta\eta =
  i k
  e^{-i\pi a} (d-1)^{a-1} \,
  \frac{\Gamma(a,-2i\omega\rstar\tau_3)}{(-2i\omega \tau_3)^a} \,
  \tau_1 \bar\eta
  - i \m
  e^{i\pi a} (d-1)^{-a} \,
  \frac{\Gamma(1{-}a,-2i\omega\rstar\tau_3)}{(-2i\omega \tau_3)^{1-a}} \,
  \tau_2 \bar\eta ,
\label {eq:deltaeta}
\end {equation}
where
$a$ is defined by (\ref{eq:a}) and
$\Gamma(a,z)$ is the incomplete $\Gamma$ function defined by
\begin {equation}
  \Gamma(\alpha,z) = \int_z^\infty dt\> t^{\alpha-1} e^{-t} .
\label {eq:Gamma}
\end {equation}
Now split $\bar\eta$ into $\bar\eta^{(+)}+\bar\eta^{(-)}$ and note that
$\tau_1\bar\eta^{(\pm)}$ and $\tau_2\bar\eta^{(\pm)}$ have $\tau_3 = \mp$.
So
\begin {multline}
  \delta\eta =
  i k e^{-i\pi a} (d-1)^{a-1}
  \left[
    \frac{\Gamma(a,-2i\omega\rstar)}{(-2i\omega)^a}\, \tau_1 \bar\eta^{(-)}
    +
    \frac{\Gamma(a,2i\omega\rstar)}{(2i\omega)^a}\, \tau_1 \bar\eta^{(+)}
  \right]
\\
  - i \m e^{i\pi a} (d-1)^{-a}
  \left[
    \frac{\Gamma(1{-}a,-2i\omega\rstar)}{(-2i\omega)^{1-a}}\,
    \tau_2 \bar\eta^{(-)}
    +
    \frac{\Gamma(1{-}a,2i\omega\rstar)}{(2i\omega)^{1-a}}\,
    \tau_2 \bar\eta^{(+)}
  \right] .
\label {eq:deta}
\end {multline}
In order to match these solutions to WKB expressions along the
Stokes lines, we need the asymptotic expansions for
$|\omega \rstar| \gg 1$.
The usual formula for the asymptotic expansion of the incomplete
Gamma function is
\begin {equation}
   \Gamma(\alpha,z) \simeq z^{\alpha-1} e^{-z}
   \qquad
   \mbox{($|z| \to \infty$ with $|\arg z| < \frac{3\pi}{2}$)} ,
\label {eq:Gammax}
\end {equation}
giving, for example,
\begin {equation}
   \Gamma(a,\pm2i\omega\rstar) \simeq
   (\pm 2i\omega \rstar)^{a-1} e^{\pm2i\omega\rstar}
\end {equation}
and
\begin {equation}
   \Gamma(1{-}a,\pm2i\omega\rstar) \simeq
   (\pm 2i\omega \rstar)^{-a} e^{\pm2i\omega\rstar}
\end {equation}
along the positive $\omega\rstar$ Stokes line.
Since $a{-}1$ and $-a$ are negative,
these factors and
the corresponding $\delta\eta$ (\ref{eq:deta}) vanish as $\omega\rstar$
becomes large, and so $\eta \to \bar\eta$.
If the same were true along the negative $\omega\rstar$ line, then the
matching near the singularity would be trivial, with
$\eta_{\omega\rstar>0} = \eta_{\omega\rstar<0}$ in (\ref{eq:WKBposneg}).
If that were true, there could be no solution satisfying the boundary
conditions required of a quasinormal mode.

The loophole is that the formula (\ref{eq:Gammax}) does not apply to all
the terms of (\ref{eq:deta}) along the negative $\omega\rstar$ Stokes line.
The phase change in $r$ in moving from the positive to the
negative $\omega\rstar$
Stokes line in fig.\ \ref{fig:stokes} is $e^{i\pi/(d-1)}$, corresponding
to a phase change in $\omega\rstar$ of $e^{i\pi}$.
The phase of $2i\omega\rstar$ is then $e^{i3\pi/2}$,
which violates the condition on the argument $z$ in
(\ref{eq:Gammax}) for the case of $\Gamma(\alpha,2i\omega\rstar)$.
We can bring the argument back within the range of validity if
we use the monodromy of the incomplete $\Gamma$ function, which is
\begin {equation}
   \Gamma(\alpha, e^{i2\pi n}z) =
   [1-e^{i2\pi n\alpha}] \Gamma(\alpha) + e^{i 2\pi n\alpha} \Gamma(\alpha,z) .
\end {equation}
For our purposes, the differences between the
positive $\omega\rstar$ and negative $\omega\rstar$ formulas for
the asymptotic expansion can
be usefully summarized as%
\footnote{
  In slightly more detail, rewrite the case of negative $\omega\rstar$
  as $\omega\rstar = e^{i\pi} \omega y$ where $\omega y$ is positive.
  Then, in this case,
  $\Gamma(\alpha, 2 i \omega\rstar)
   = \Gamma(\alpha, 2 e^{3i\pi/2} \omega y)
   = [1-e^{i2\pi\alpha}]\Gamma(\alpha)
     + e^{i2\pi\alpha}\Gamma(\alpha,2 e^{-i\pi/2} \omega y)
   \simeq [1-e^{i2\pi\alpha}]\Gamma(\alpha)
     + e^{i2\pi\alpha}(2 e^{-i\pi/2} \omega y)^{\alpha-1} e^{-2i\omega y}
   = [1-e^{i2\pi\alpha}]\Gamma(\alpha)
     + (2 e^{3i\pi/2} \omega y)^{\alpha-1} e^{-2i\omega y}
   = [1-e^{i2\pi\alpha}]\Gamma(\alpha)
     + (2 i\omega\rstar)^{\alpha-1} e^{2i\omega\rstar}
  $.
  The second term in the last expression is the same asymptotic
  formula that we would get for
  $\Gamma(\alpha, 2i\omega\rstar)$
  in the case of positive $\omega\rstar$.
}
\begin {equation}
  \Gamma(\alpha,\mp 2i\omega\rstar)
  \to [\mbox{positive $\omega\rstar$ formula}]
      + (1-e^{i2\pi\alpha}) \, \Gamma(\alpha) \,
          \delta_{\mp,+} \,\theta(-\omega\rstar) ,
\label {eq:GammaMonodromy}
\end {equation}
where $\theta$ is the step function.
Applying (\ref{eq:GammaMonodromy}) to (\ref{eq:deta}),
and remembering that $\eta = \bar\eta + \delta\eta$, gives us
the relationship between the WKB solutions (\ref{eq:WKBposneg}) along the
positive and negative $\omega\rstar$ Stokes lines (to first order in
$k$ and $m$):
\begin {multline}
  \eta_{\omega\rstar < 0} =
  \eta_{\omega\rstar > 0}
  + i k e^{-i\pi a} (d-1)^{a-1} \,
       \frac{(1-e^{i 2\pi a}) \, \Gamma(a)}{(2i\omega)^a} \,
  \tau_1 \eta^{(+)}_{\omega\rstar > 0}
\\
  - i \m e^{i\pi a} (d-1)^{-a} \,
       \frac{(1-e^{i 2\pi(1-a)}) \, \Gamma(1-a)}{(2i\omega)^{1-a}} \,
  \tau_2 \eta^{(+)}_{\omega\rstar > 0}
  .
\label {eq:etamatch}
\end {multline}
This may be rewritten in the form
\begin {equation}
  \eta_{\omega\rstar < 0} =
  \eta_{\omega\rstar > 0}
  + \left(
      \tau_1 \lambda \frac{(k/\pi T)}{(\omega/\pi T)^a}
    + \tau_2 \Lambda \frac{\m}{(\omega/\pi T)^{1-a}} \,
    \right) \eta^{(+)}_{\omega\rstar > 0}
\label {eq:etamatch2}
\end {equation}
where $\lambda$ and $\Lambda$ are defined as in (\ref{eq:lLambda}),
and the temperature $T$ is given by (\ref{eq:T}).

% ----------------------------------------------------------------------------

\subsection{Behavior near the boundary}

Near the boundary ($z=0$), the SAdS Dirac equation becomes that of pure
AdS, corresponding to replacing $f$ by 1 in (\ref{eq:diracpsi}).
In our basis (\ref{eq:basis}), the solution which vanishes at the
boundary is
\begin {equation}
   \psi \propto
   \sqrt{\pi\Omega z}
   \begin{pmatrix}
     J_{\m-\tfrac12}(\Omega z)
        + i \, \frac{\Omega}{\Omega-k} \, J_{\m+\tfrac12}(\Omega z) \\
     \frac{\Omega}{\Omega-k} \, J_{\m+\tfrac12}(\Omega z)
        + i J_{\m-\tfrac12}(\Omega z)
   \end{pmatrix}
   ,
\end {equation}
where $\Omega^2 \equiv \omega^2 - |\k|^2$.
We're interested in the limit of large $|\omega| \gg |\k|$.
We're also interested in the asymptotic expansion of this solution
away from the boundary (i.e. $\omega z \gg 1$) in order to match
to the WKB solution (\ref{eq:WKBpos}) along the positive $\omega\rstar$
Stokes line.  These further approximations then give
\begin {equation}
   \psi \propto
   \begin{pmatrix}
     - i e^{-i\m\pi/2} e^{i\omega z} \\
     e^{i\m\pi/2} e^{-i\omega z}
   \end {pmatrix}
   .
\end {equation}
For small $z$, the relationship between $z$ and the tortoise
coordinate (\ref{eq:rstar}) is
\begin {equation}
   z \simeq \rstaro - \rstar ,
\end {equation}
where
\begin {equation}
   \rstaro \equiv \rstar \bigl|_{\rm boundary}
   = \int_0^\infty \frac{dz}{f}
   = \frac{e^{i\pi/d}}{4T\sin(\pi/d)}
   \,,
\label {eq:rstaro}
\end {equation}
and so
\begin {equation}
   \psi \propto
   e^{-i\omega\rstar\tau_3}
   \begin{pmatrix}
     - i e^{-i\m\pi/2} e^{i\omega\rstaro} \\
     e^{i\m\pi/2} e^{-i\omega\rstaro}
   \end {pmatrix}
   .
\label {eq:psibdy}
\end {equation}
Comparing to (\ref{eq:WKBpos}), we identify
\begin {equation}
   \eta_{\omega\rstar>0} =
   C \begin{pmatrix}
     - i e^{-i\m\pi/2} e^{i\omega\rstaro} \\
     e^{i\m\pi/2} e^{-i\omega\rstaro}
   \end {pmatrix}
   \equiv
   \begin {pmatrix}
      \eta^+_{\omega\rstar>0} \\[4pt]
      \eta^-_{\omega\rstar>0}
   \end {pmatrix}
\label {eq:etabdy}
\end {equation}
for some proportionality constant $C$.
This decomposes into
\begin {equation}
   \eta^{(+)}_{\omega\rstar>0} =
   \begin{pmatrix}
     \eta^+_{\omega\rstar>0} \\[4pt]
     0
   \end {pmatrix}
   \qquad \mbox{and} \qquad
   \eta^{(-)}_{\omega\rstar>0} =
   \begin {pmatrix}
     0 \\[4pt]
     \eta^-_{\omega\rstar>0}
   \end {pmatrix} .
\end {equation}

% ---------------------------------------------------------------------------

\subsection{Putting it together}

The horizon condition for the quasi-normal modes is that
$\eta^{(-)}_{\omega\rstar < 0}$ vanish, i.e. that there is no
$e^{+i\omega\rstar}$ component of the WKB solution along the
negative $\omega\rstar$ Stokes line of fig.\ \ref{fig:stokes}.
From (\ref{eq:etamatch2}), this condition gives
\begin {equation}
  0 =
  \eta^{(-)}_{\omega\rstar > 0}
  + \left(
      \tau_1 \lambda \frac{(k/\pi T)}{(\omega/\pi T)^a}
    + \tau_2 \Lambda \frac{\m}{(\omega/\pi T)^{1-a}} \,
    \right) \eta^{(+)}_{\omega\rstar > 0}
\end {equation}
for the WKB coefficients on the positive $\omega\rstar$ line.
In terms of components,
\begin {equation}
  0 =
  \eta^-_{\omega\rstar > 0}
  + \left(
      \lambda \frac{(k/\pi T)}{(\omega/\pi T)^a}
    + i \Lambda \frac{\m}{(\omega/\pi T)^{1-a}} \,
    \right) \eta^+_{\omega\rstar > 0} .
\end {equation}
Using the explicit expression (\ref{eq:etabdy}) derived from matching
the solution along that line to the desired behavior at the
boundary $z{=}0$ then gives
\begin {equation}
  \left(
      i \lambda \frac{(k/\pi T)}{(\omega/\pi T)^a}
      - \Lambda \frac{\m}{(\omega/\pi T)^{1-a}} \,
  \right)
  e^{-i\m\pi} e^{2i\omega\rstaro}
  = 1 .
\label {eq:condition0}
\end {equation}
This condition is satisfied when
\begin {equation}
  \ln\left(
      i \lambda \frac{(k/\pi T)}{(\omega/\pi T)^a}
      - \Lambda \frac{\m}{(\omega/\pi T)^{1-a}} \,
  \right)
  -i\m\pi + 2i\omega\rstaro
  = 2i n\pi
\label {eq:condition}
\end {equation}
for some integer $n$.  Treating $n$ as large, solving for
$\omega$ through $O(n^0)$, and using the explicit expression
(\ref{eq:rstaro}) for $\rstaro$, gives the result
(\ref{eq:dirac}) previewed in the introduction.

% ---------------------------------------------------------------------------

\subsection{Range of validity}

The perturbative treatment of $k$ and $m$ near the singularity
assumed that the $k$ and $m$ mixing coefficients in (\ref{eq:etamatch2})
were small.  Parametrically, that's
\begin {equation}
   \left| \frac{k/T}{(\omega/T)^a} \right| \ll 1
   \qquad \mbox{and} \qquad
   \left| \frac{\m}{(\omega/\pi T)^{1-a}} \right| \ll 1 .
\end {equation}
In terms of the parameter $n \sim |\omega_n|/T$ of our asymptotic
formula (\ref{eq:dirac}), those conditions are equivalent to
\begin {equation}
   n \gg \mbox{both}~
   \left( \frac{|\k|}{T} \right)^{1/a}
   ~\mbox{and}~
   \m^{1/(1-a)} ,
\label {eq:ncondition1}
\end {equation}
which, for example,
for $D{=}5$ would be $n \gg (|\k|/T)^3$ and $\m^{3/2}$.
(We will assume for simplicity that $\m \gtrsim 1$, which is
true of fermions coming from compactification of Type II SUGRA
on the $S^5$ of (S)AdS$_5 \times S^5$.)
In the earlier paper \cite{dirac} on spin-$\tfrac12$ quasinormal
modes, we did not include the mass term in the argument of the logarithm
in our result (\ref{eq:dirac}).  The assumption that the $k$ term dominates
the $m$ term in the argument of the logarithm requires the
additional condition that
\begin {equation}
   n \gg \left( \frac{\m}{|\k|/T} \right)^{1/(1-2a)} ,
\end {equation}
which for $D{=}5$ is $n \gg (\m T/|\k|)^3$.
Including the mass term in the logarithm therefore extends the
range of validity of our result in those cases where
the second condition in (\ref{eq:ncondition1}) is more important
than the first---that is, in cases where
\begin {equation}
   \frac{|\k|}{T} \ll \m^{a/(1-a)} = \m^{(d-2)/d}
\end {equation}
(e.g.\ $|\k|/T \ll \m^{1/2}$ in $D{=}5$).

% ============================================================================

\section {Spin \boldmath$\tfrac32$}
\label {sec:spin32}

\subsection {Basic equations}

We now turn to the spin-$\tfrac32$ equation of motion (\ref{eq:spin32eq}),
\begin {equation}
   \Gamma^{MNP} D_N \Psi_P + m \Gamma^{MN} \Psi_N = 0 .
\label {eq:spin32eq2}
\end {equation}
If $m$ is not the gravitino mass (\ref{eq:mgravitino}), one can show
that, in an Einstein spacetime (such as SAdS), this equation implies that%
\footnote{
  See, for example, the discussion in ref.\ \cite{Policastro}.
}
\begin {equation}
   \Gamma^M \Psi_M = 0
   \qquad \mbox{and} \qquad
   D^M \Psi_M = 0 .
\label {eq:PsiMconditions}
\end {equation}
If $m$ {\it is} the gravitino mass, then there is a gauge symmetry
(\ref{eq:gauge}), and one may (i) choose $\Gamma^M \Psi_M = 0$ as
a gauge condition and (ii) show that the equation of motion then
implies $D^M \Psi_M = 0$.  In consequence, we will assume
(\ref{eq:PsiMconditions}) in all cases.
These conditions allow one to rewrite the equation of motion
(\ref{eq:spin32eq2}) as
\begin {equation}
   \Gamma^M D_M \Psi_N - m \Psi_N = 0 ,
\label {eq:spin32simple}
\end {equation}
supplemented by the constraint that one only keep solutions
to (\ref{eq:spin32simple}) that have $\Gamma^M \Psi_M = 0$.%
\footnote{
   $D^M \Psi_M = 0$ can be shown to follow for any solution
   of (\ref{eq:spin32simple}) that satisfies $\Gamma^M \Psi_M = 0$.
}
Note that (\ref{eq:spin32simple})
differs from the Dirac equation (\ref{eq:diraceq})
because the covariant derivative in (\ref{eq:spin32simple})
contains a Christoffel term that operates
on the vector index $N$ of $\Psi_N$.

There are any number of different conventions one might now use
to rescale $\Psi$ in order to write out convenient explicit equations
in terms of components.  We choose the following generalization
of the rescaling (\ref{eq:PsipsiDirac}) that we used in
the Dirac case:
\begin {equation}
   \Psi_M \equiv (-g g^{zz})^{-1/4} e^m_M \psi_m
   \propto z^{d/2} f^{-1/4} e^m_M \psi_m ,
\end {equation}
where $M$ is a curved-space index, $m$ is a flat-space index, and
$e^m_M = \sqrt{|g^{mm}|}\,\delta^m_M$ (no sum) is the inverse vielbein.
With this rewriting, the explicit equations of motion are%
\footnote{
  One can get the same equations from (25--27) of Policastro \cite{Policastro}
  by (i) switching coordinates from his $u$ to our $z$, which are related by
  $u = z^2$ (in our working units, where $\zh=1$);
  (ii) rewriting his
  $\psi_n$ as our $z^{d/2} f^{-1/4} \psi_n$ (with $d{=}4$)
  and (iii) switching the mass sign convention, replacing his $mR$ by our
  $-\m$.
  Note that there is no transformation associated with the subscript on
  $\psi_m$ in going from $u$ to $z$ because it is a flat-space index.
}
\begin {subequations}
\label {eq:eom}
\begin {align}
   &
   {\cal D}_{1/2} \psit
   - \left( \frac{f}{z} - \frac{f'}{2}\right) \gammaz\gammat \psiz = 0 ,
\label {eq:eom0}
\\
   &
   {\cal D}_{1/2} \psiz
   + \frac{f}{z} \psiz
   + \frac{f'}{2} \, \gammaz\gammat \psit = 0 ,
\label {eq:eom5}
\\
\intertext{and}
   &
   {\cal D}_{1/2} \psi_i
   - \frac{f}{z} \, \gamma^5\gamma^i \psiz = 0 .
\label {eq:eomi}
\end {align}
\end {subequations}
Here and throughout, $i = 1,\cdots,d-1$ runs over the spatial dimensions
parallel to the boundary, ``$\psi_5$'' is our generic notation
for the component $\psi_z$,
and ${\cal D}_{1/2}$ is the same differential operator
(\ref{eq:D1/2}) as in the spin-$\tfrac12$ case.

We are only interested in solutions to (\ref{eq:eom})
which satisfy the
constraint $\Gamma^M \Psi_M = 0$, which is equivalent to
$\gamma^m \psi_m = 0$.
We may use this to determine
$\k\cdot{\bm\psi}$ in terms of $\Psi_0$ and $\Psi_5$ as
\begin {align}
   \k\cdot{\bm\psi} &=
   - \frac{i}{2f^{1/2}}
   \left[
       \left( \gammaz {\cal D}_{1/2} - 2 \m \frac{f^{1/2}}{z} \right)
       (-\gamma^i \psi_i)
     - (d-1)\frac{f}{z} \, \psiz
   \right]
\nonumber\\
   &=
   - \frac{i}{2f^{1/2}}
   \left[
       \left( \gammaz {\cal D}_{1/2} - 2 \m \frac{f^{1/2}}{z} \right)
       (\gammat \psit + \gammaz \psiz)
     - (d-1)\frac{f}{z} \, \psiz
   \right]
   ,
\label {eq:psi3}
\end {align}
where the first equality follows from the equation of motion
(\ref{eq:eomi}) and the explicit form of ${\cal D}_{1/2}$,
and the second equality follows from the constraint $\gamma^m\psi_m=0$.

\subsection{``Transverse'' solutions}

There are two types of solutions to these equations satisfying the
constraint.
We describe here the first type, which are characterized
by $\psiz=0$.
We will call such solutions ``transverse'' polarizations.
As noted previously by others \cite{Policastro,Erdmenger}, and as we review
below, these modes decouple and satisfy a simple Dirac equation.
We will see below that they exist only for $D > 4$.

From $\psiz{=}0$, it follows by (\ref{eq:eom5}) that $\psit{=}0$
and thence by (\ref{eq:psi3}) that $\k\cdot{\bm\psi} = 0$ as well.
That means that $\psi_m$ is zero except for the subset of
indices $m$ corresponding to the $D{-}3$ spatial directions parallel to
the boundary but transverse to $\k$.  For simplicity of
presentation, we will focus on the explicit case $D{=}5$ but will
state the generalization to larger $D$ at the end.
For $D{=}5$, take $\k$ to point in the $x^3$ direction.
Then, for the transverse solutions under consideration,
$\psi_m$ is zero except for $\psi_1$ and $\psi_2$.  By
the $\gamma$-traceless condition, they must be related by
\begin {equation}
   \psi_2 = - \gamma^2 \gamma^1 \psi_1 .
\label {eq:psi2}
\end {equation}
Since $\psiz = 0$, the equation of motion (\ref{eq:eomi}) for both
$\psi_1$ and $\psi_2$
is simply the Dirac equation (\ref{eq:diracpsi}),
\begin {equation}
   {\cal D}_{1/2} \psi_i = 0 .
\end {equation}
So let $\psi_1$ be any solution to the Dirac equation.  Since
$[{\cal D}_{1/2},\gamma^2\gamma^1] = 0$, then $\psi_2$ given by
(\ref{eq:psi2}) will also solve the Dirac equation, and so
we have a consistent solution.  The quasinormal mode frequencies
of these solutions will simply be those of the Dirac equation.

For $D>5$, we can similarly construct transverse solutions by taking only
two of the $\psi_m$'s non-zero, with those two $m$'s taken from the
transverse spatial directions $1,2,\cdots,D{-}3$.  One choice
of a complete basis of all transverse solutions is to use
$\psi_1$ and $\psi_{i_\perp}$, with $i_\perp$ selected from the
$D{-}4$ choices
$2,3,\cdots,D{-}3$.  The quasinormal mode frequencies of transverse
solutions are then simply those of the Dirac equation, but
with a degeneracy of $D{-}4$ compared to a single Dirac field.

\subsection{Setup for finding non-transverse solutions}

With the transverse modes out of the way, we will focus
exclusively on non-transverse
($\psiz\not=0$) modes in what follows.
Because $\psit$ and $\psiz$ by themselves satisfy a closed set of
equations of motion (\ref{eq:eom0},b),
we may ignore what
the other components $\psi_i$ are doing if our goal is just to
find the quasinormal frequencies $\omega_n$.%
\footnote{
   This argument leaves the question
   of whether all such solutions for $\psit$ and
   $\psiz$ may be extended to solutions for the other components $\psi_i$
   such that the constraint $\gamma^m\psi_m = 0$ is satisfied.
   Consider $D{=}5$ for concreteness, but the argument generalizes
   to all $D>3$.
   Enforcing $\gamma^m \psi_m = 0$, (\ref{eq:psi3}) may be used to
   explicitly determine
   $\gamma^{i_\perp} \psi_{i_\perp} = \gamma^1 \psi_1 + \gamma^2 \psi_2$ in terms
   of $\psi_0$ and $\psi_5$.  One may then verify that this
   result for $\gamma^1 \psi_1 + \gamma^2 \psi_2$ is
   compatible with the equations of motion for $\psi_1$ and $\psi_2$,
   and so all is well.
   In the quasinormal mode problem, the transverse and non-transverse
   modes have different frequencies, and so
   $\psi_1$ and $\psi_2$ may individually be uniquely determined
   by combining the result for $\gamma^1 \psi_1 + \gamma^2 \psi_2$
   with the condition $\gamma^1 \psi_1 - \gamma^2 \psi_2 = 0$ that
   no transverse component is present.
}

The naive large-$\omega$ limit of
the $(\psi_0,\psi_5)$ equations of motion
(\ref{eq:eom0},b) is
\begin {align}
   &
   [-f\partial_z + i\omega\tau_3] \psit
   - \left( \frac{f}{z} - \frac{f'}{2}\right) \tau_3 \psiz \simeq 0 ,
\\
   &
   [-f\partial_z + i\omega\tau_3] \psiz
   + \frac{f}{z} \, \psiz
   + \frac{f'}{2} \, \tau_3 \psit = 0 ,
\end {align}
which is the analog to eq.\ (\ref{eq:naive}) from the spin-$\tfrac12$ analysis.
The corresponding solution is
\begin {equation}
   \begin{pmatrix} \psi_0 \\ \psi_5 \end{pmatrix} =
   \omega z f^{-1/2} e^{-i\omega\rstar\tau_3}
   \begin {pmatrix}
     \tau_3 \left[
       -\eta
       + \chi \left( \frac{2f}{z} - \int \frac{f'}{z}\right)
     \right] \\
     \eta
     + \chi \int \frac{f'}{z}
   \end {pmatrix} ,
\label {eq:WKB}
\end {equation}
where $\eta$ and $\chi$ are arbitrary constant spinors.%
\footnote{
  So far, we have not specialized to units where $\zh=1$, and so
  it may be useful in what follows
  to note that our $\eta$ and $\chi$ have different
  units: $[\chi] = [\eta] \times$length.
}
By convention,
we will choose the lower limit on the integrals in (\ref{eq:WKB})
to be zero, i.e.
\begin {equation}
   \int \frac{f'}{z} \equiv \int_0^z dz_1 \> \frac{f'(z_1)}{z_1} \,.
\end {equation}

Similar to our treatment of the spin-$\tfrac12$ case, we now promote
$\eta$ and $\chi$ to functions of $z$.
Plugging (\ref{eq:WKB}) into (\ref{eq:eom0},b) then recasts the
exact equations of motion to
\begin {subequations}
\begin {align}
   \partial_{\rstar} \eta
   - \left( \frac{2f}{z} - \int\frac{f'}{z} \right) \partial_{\rstar}\chi
   &=
   f^{1/2} e^{2i\omega\rstar\tau_3}
   \left( k \tau_1 + \frac{\m}{z} \, \tau_2 \right)
   \left[ \eta - \chi \left( \frac{2f}{z} - \int\frac{f'}{z} \right) \right],
\\
   \partial_{\rstar} \eta
   + \left( \int \frac{f'}{z} \right) \partial_{\rstar}\chi
   &=
   f^{1/2} e^{2i\omega\rstar\tau_3}
   \left( k \tau_1 + \frac{\m}{z} \, \tau_2 \right)
   \left[ - \eta - \chi \int\frac{f'}{z} \right] ,
\end {align}
\end {subequations}
which may also be written as
\begin {subequations}
\label {eq:etaxi1}
\begin {align}
   \partial_{\rstar} \chi
   &=
   - f^{1/2} e^{2i\omega\rstar\tau_3}
   \left( k \tau_1 + \frac{\m}{z} \, \tau_2 \right)
   \frac{z}{f}
   \left[ \eta - \chi \left( \frac{f}{z} - \int\frac{f'}{z} \right) \right],
\\
   \partial_{\rstar} \eta
   &=
   - \left( \int \frac{f'}{z} \right) \partial_{\rstar}\chi
   - f^{1/2} e^{2i\omega\rstar\tau_3}
   \left( k \tau_1 + \frac{\m}{z} \, \tau_2 \right)
   \left[ \eta + \chi \int\frac{f'}{z} \right] .
\end {align}
\end {subequations}
% ----------------------------------------------------------------------------

\subsection{Behavior near the singularity}

Near the singularity, we approximate
$f$ by $-z^d$ as in the spin-$\tfrac12$ case.  Then (\ref{eq:etaxi1})
becomes
\begin {align}
   \partial_{\rstar} \chi
   &\simeq
   - i
   e^{2i\omega\rstar\tau_3}
   \left( k \tau_1 + \frac{\m}{z}\,\tau_2 \right)
   \left[ z^{1-\frac{d}{2}} \eta - \frac{1}{(d-1)} \, z^{\frac{d}{2}} \chi \right],
\\
   \partial_{\rstar} \eta
   &\simeq
   - i e^{2i\omega\rstar\tau_3}
   \left( k \tau_1 + \frac{\m}{z}\,\tau_2 \right)
   \left[ \frac{1}{(d-1)} \, z^{\frac{d}{2}} \eta
          + \frac{d(d-2)}{(d-1)^2} \, z^{\frac{3d}{2}-1} \chi \right] .
\end {align}
Now treat $k$ and $\m$ as perturbations; take
\begin {equation}
   \eta(z) = \bar\eta + \delta\eta(z) ,
   \qquad
   \chi(z) = \bar\chi + \delta\chi(z) ;
\end {equation}
expand to first order in small quantities; and integrate the
resulting equations for $\delta\eta$ and $\delta\chi$ to get
\begin {align}
   \delta\chi
   &=
   i
   \int_{\rstar}^\infty d\rstar \>
   e^{2i\omega\rstar\tau_3}
   \left( k \tau_1 + \frac{\m}{z}\,\tau_2 \right)
   \left[ z^{1-\frac{d}{2}} \bar\eta
          - \frac{1}{(d-1)} \, z^{\frac{d}{2}} \bar\chi \right],
\\
   \delta\eta
   &=
   i
   \int_{\rstar}^\infty d\rstar \>
   e^{2i\omega\rstar\tau_3}
   \left( k \tau_1 + \frac{\m}{z}\,\tau_2 \right)
   \left[ \frac{1}{(d-1)} \, z^{\frac{d}{2}} \bar\eta
          + \frac{d(d-2)}{(d-1)^2} \, z^{\frac{3d}{2}-1} \bar\chi \right] .
\end {align}
Using (\ref{eq:a}), (\ref{eq:zrstar}), and (\ref{eq:Gamma}),
the integration gives
\begin {align}
   \delta\chi
   &=
   i k e^{-i\pi a} (d-1)^a
   \left[
         \frac{\Gamma(a{+}1,-2i\omega\rstar\tau_3)}
              {(-2i\omega\tau_3)^{a+1}}
         \tau_1\bar\eta
      +
         \frac{\Gamma(a,-2i\omega\rstar\tau_3)}
              {(d-1)^2(-2i\omega\tau_3)^a}
         \tau_1\bar\chi
  \right]
\nonumber\\
  &+
   i \m e^{i\pi a} (d-1)^{1-a}
   \left[
      -
         \frac{\Gamma(2{-}a,-2i\omega\rstar\tau_3)}
              {(-2i\omega\tau_3)^{2-a}}
         \tau_2\bar\eta
      -
         \frac{\Gamma(1{-}a,-2i\omega\rstar\tau_3)}
              {(d-1)^2 (-2i\omega\tau_3)^{1-a}}
         \tau_2\bar\chi
  \right] ,
\label {eq:chistuff}
\\
   \delta\eta
   &=
   i k e^{-i\pi a} (d-1)^a
   \biggl[
      -
         \frac{\Gamma(a,-2i\omega\rstar\tau_3)}
              {(d-1)^2 (-2i\omega\tau_3)^a}
         \tau_1\bar\eta
%\nonumber\\ & \hspace{12em}
      +
         \frac{d(d-2)\,\Gamma(a{-}1,-2i\omega\rstar\tau_3)}
              {(d-1)^4 (-2i\omega\tau_3)^{a-1}}
         \tau_1\bar\chi
  \biggr]
\nonumber\\
  &+
   i \m e^{i\pi a} (d-1)^{1-a}
   \biggl[
         \frac{\Gamma(1{-}a,-2i\omega\rstar\tau_3)}
              {(d-1)^2 (-2i\omega\tau_3)^{1-a}}
         \tau_2\bar\eta
%\nonumber\\ & \hspace{12em}
      -
         \frac{d(d-2)\,\Gamma(-a,-2i\omega\rstar\tau_3)}
              {(d-1)^4 (-2i\omega\tau_3)^{-a}}
         \tau_2\bar\chi
  \biggr] .
\end {align}
Using the monodromy relation (\ref{eq:GammaMonodromy}),%
\footnote{
  There is a slight difference between the situation here and
  in the spin-$\tfrac12$ case.
  In the spin-$\tfrac12$ case, the result (\ref{eq:deta}) for
  $\delta\eta$ approached zero in the limit of large positive
  $\omega\rstar$.  This is not true for the
  $\Gamma(a{+}1,\cdots)$ and $\Gamma(2{-}a,\cdots)$ terms in
  (\ref{eq:chistuff}) because $a{+}1$ and $2{-}a$ exceed $1$
  [see (\ref{eq:Gammax})].
  These terms (along with the others) account for the $O(k/\omega)$
  and $O(\m/\omega)$
  corrections that were ignored in the WKB solution (\ref{eq:WKB}).
  We could tediously keep track of those corrections, and check the matching
  of (\ref{eq:chistuff}) to an improved WKB solution, but this is
  unnecessary since all we care about is the {\it difference}\/
  of the asymptotic formulas along the positive and negative
  $\omega\rstar$ lines, which is captured by (\ref{eq:diff}).
}
\begin {align}
   \begin{pmatrix} \chi \\ \eta \end{pmatrix}_{\omega\rstar<0}
   &=
   \begin{pmatrix} \chi \\ \eta \end{pmatrix}_{\omega\rstar>0}
\nonumber\\ &
   +i k e^{-i\pi a} (1-e^{i2\pi a}) (d-1)^a
   \begin{pmatrix}
         \frac{\Gamma(a)}
              {(d-1)^2(2i\omega)^a}
      &
         \frac{\Gamma(a+1)}
              {(2i\omega)^{a+1}}
      \\[6pt]
         \frac{d(d-2)\,\Gamma(a-1)}
              {(d-1)^4 (2i\omega)^{a-1}}
     &
        -\frac{\Gamma(a)}
              {(d-1)^2 (2i\omega)^a}
  \end{pmatrix}
  \begin{pmatrix} \tau_1 \chi^{(+)} \\ \tau_1 \eta^{(+)} \end{pmatrix}_{\omega\rstar>0}
\nonumber\\ &
   +i \m e^{i\pi a} (1-e^{i2\pi a}) (d-1)^{1-a}
   \begin{pmatrix}
         -\frac{\Gamma(1-a)}
               {(d-1)^2(2i\omega)^{1-a}}
      &
         -\frac{\Gamma(2-a)}
               {(2i\omega)^{2-a}}
      \\[6pt]
         -\frac{d(d-2)\,\Gamma(-a)}
               {(d-1)^4 (2i\omega)^{-a}}
     &
         \frac{\Gamma(1-a)}
              {(d-1)^2 (2i\omega)^{1-a}}
  \end{pmatrix}
  \begin{pmatrix} \tau_2 \chi^{(+)} \\ \tau_2 \eta^{(+)} \end{pmatrix}_{\omega\rstar>0}
  .
\label {eq:diff}
\end {align}
This can be rewritten as
\begin {equation}
   \begin{pmatrix} \chi \\ \eta \end{pmatrix}_{\omega\rstar<0}
   =
   \begin{pmatrix} \chi \\ \eta \end{pmatrix}_{\omega\rstar>0} +
   \left[
      \tau_1 \lambda \, \frac{k/\pi T}{(\omega/\pi T)^a}
      \utilde{\cal M}
    + \tau_2 \Lambda \, \frac{\m}{(\omega/\pi T)^{1-a}}
      \utilde{\cal N}
   \right]
   \begin{pmatrix} \chi^{(+)} \\ \eta^{(+)} \end{pmatrix}_{\omega\rstar>0}
\end {equation}
with $\lambda$ and $\Lambda$ defined as in (\ref{eq:lLambda}),
\begin {equation}
   \utilde{\cal M} \equiv
   \begin{pmatrix}
      \frac{1}{(d-1)}
      & \frac{(d-1)a}{2i\omega}
      \\[4pt]
      \frac{d(d-2) 2i\omega}{(d-1)^{3} (a-1)}
      &
      -\frac{1}{(d-1)}
   \end{pmatrix} ,
\end {equation}
and
\begin {equation}
   \utilde{\cal N} \equiv
   \begin{pmatrix}
      \frac{1}{(d-1)}
      & \frac{(d-1)(1-a)}{2i\omega}
      \\[4pt]
      \frac{d(d-2) 2i\omega}{(d-1)^{3} (-a)}
      &
      -\frac{1}{(d-1)}
   \end{pmatrix} .
\end {equation}
In general, we will use undertildes to denote $2\times2$ matrices
that act on the space of $(\chi,\eta)$.

It's now convenient to simplify using the definition (\ref{eq:a}) of
$a$ and furthermore to rescale the field $\chi$ as%
\footnote{
  Note that $\hat\chi$, unlike $\chi$,
  has the same engineering dimension as $\eta$.
}
\begin {equation}
   \chi = \frac{(d-1)}{4i\omega} \, \hat\chi ,
\label {eq:chihat}
\end {equation}
giving
\begin {equation}
   \begin{pmatrix} \hat\chi \\ \eta \end{pmatrix}_{\omega\rstar<0}
   =
   \begin{pmatrix} \hat\chi \\ \eta \end{pmatrix}_{\omega\rstar>0} +
   \left[
      \tau_1 \lambda \, \frac{k/\pi T}{(\omega/\pi T)^a}
      \utilde{\hat{\cal M}}
    + \tau_2 \Lambda \, \frac{\m}{(\omega/\pi T)^{1-a}}
      \utilde{\hat{\cal N}}
   \right]
   \begin{pmatrix} \hat\chi^{(+)} \\ \eta^{(+)} \end{pmatrix}_{\omega\rstar>0}
\label {eq:etaxi3}
\end {equation}
with
\begin {equation}
   \utilde{\hat{\cal M}} =
   \frac{1}{d-1}
   \begin{pmatrix}
      1
      & +(d-2)
      \\[4pt]
      -(d-2)
      &
      -1
  \end{pmatrix}
  ,
  \qquad
   \utilde{\hat{\cal N}} =
   \frac{1}{d-1}
   \begin{pmatrix}
      1
      & +d
      \\[4pt]
      -d
      &
      -1
  \end{pmatrix}
  .
\label {eq:MNhat}
\end {equation}
The only difference between (\ref{eq:etaxi3}) and the corresponding
spin-$\tfrac12$ formula (\ref{eq:etamatch2}) is the inclusion of
the matrix factors $\utilde{\hat{\cal M}}$ and $\utilde{\hat{\cal N}}$.

% ----------------------------------------------------------------------------

\subsection{Behavior near the boundary}

The desired solutions near the boundary, adapted from
ref.\ \cite{Volovich}, are
\begin {align}
  \psi_0 = &
  \pi^{1/2} (\Omega z)^{1/2} \left[
     \frac{i \gamma^\mu k_\mu}{\Omega} \, J_{\m+\frac12}(\Omega z)
     + J_{\m-\frac12}(\Omega z)
  \right] \xi_0
\nonumber\\ & \quad
  + \pi^{1/2} (\Omega z)^{3/2} \left[
     - \frac{k_0 \gamma^\mu k_\mu}{\Omega^2} \, J_{\m+\frac32}(\Omega z)
     + \frac{i k_0}{\Omega} \, J_{\m+\frac12}(\Omega z)
     + \frac{\gamma_0}{\Omega z} \, J_{\m+\frac12}(\Omega z)
  \right] \xi_5 ,
\\
  \psi_5 = &
  \pi^{1/2} (\Omega z)^{3/2} \left[
     \frac{i \gamma^\mu k_\mu}{\Omega} \, J_{\m+\frac12}(\Omega z)
     + J_{\m-\frac12}(\Omega z)
  \right] \xi_5 ,
\label {eq:boundary1}
\end {align}
where $\xi_0$ and $\xi_5$ are constant spinors that are
$\gammaz=+1$ eigenstates of $\gammaz$, and
$\Omega \equiv \sqrt{- k^\mu k_\mu}$.
Taking the large $\omega$ limit (with $\k$ fixed) replaces
$\Omega$ by $\omega$ above.  In our representation
(\ref{eq:basis}) for the $\gamma$ matrices, we may write
\begin {equation}
   \xi_I = \frac{C_I}{\sqrt2} \begin{pmatrix} -i \\ 1 \end{pmatrix}
\end {equation}
in the space that the $\tau$ matrices act on, where $C_0$
and $C_5$ are simple numbers.
Taking the asymptotic expansion $\omega z \gg 1$ of
(\ref{eq:boundary1}), one then finds
\begin {subequations}
\label{eq:psibdyWKB}
\begin {align}
  \psi_0 &\simeq
  \begin{pmatrix}
     -i(C_0 - \omega z C_5) e^{-i\m\pi/2} e^{i\omega z} \\
     (C_0 + \omega z C_5) e^{i\m\pi/2} e^{-i\omega z}
  \end {pmatrix}
  \simeq
  \omega z
  e^{-i\omega\rstar\tau_3}
  \tau_3
  \begin{pmatrix}
     \bigl(iC_5 - \frac{iC_0}{\omega z}\bigr) e^{-i\m\pi/2} e^{i\omega\rstaro} \\[4pt]
     \bigl(- C_5 - \frac{C_0}{\omega z}\bigr) e^{i\m\pi/2} e^{-i\omega\rstaro}
  \end {pmatrix} ,
\label {eq:psi0thing1}
\\
  \psi_5 &\simeq
  \begin{pmatrix}
     -i \omega z C_5 e^{-i\m\pi/2} e^{i\omega z} \\
     \omega z C_5 e^{i\m\pi/2} e^{-i\omega z}
  \end {pmatrix}
  \simeq
  \omega z
  e^{-i\omega\rstar\tau_3}
  \begin{pmatrix}
     -i C_5 e^{-i\m\pi/2} e^{i\omega\rstaro} \\
     C_5 e^{i\m\pi/2} e^{-i\omega\rstaro}
  \end {pmatrix} .
\end {align}
\end {subequations}
Note that we have been slightly inconsistent: We have kept the leading
terms for $\omega z \gg 1$ that are proportional to $C_0$ and $C_5$.
However, in (\ref{eq:psi0thing1}),
the next-order term proportional to $C_5$ would seemingly
be the same order in $1/\omega z$ as the leading term proportional
to $C_0$.  However, this will not be an issue in what follows.

We now want to match to the WKB solution (\ref{eq:WKB}),
which near the boundary becomes
\begin {subequations}
\label {eq:psibdyWKB2}
\begin {align}
  \psi_0 &=
  \omega z e^{-i\omega\rstar\tau_3}
  \tau_3 \Bigl(
       -\eta
       + \frac{2}{z} \, \chi
     \Bigr) ,
\label {eq:psi0thing2}
\\
  \psi_5 &=
  \omega z e^{-i\omega\rstar\tau_3}
  \eta .
\end {align}
\end {subequations}
Here again, we are being slightly inconsistent: Corrections to
the approximation (\ref{eq:WKB}) due to the effects of $\m$,
which were ignored in that equation, could give $O(\m/z)$
corrections to the $\eta$ term
in (\ref{eq:psi0thing2}), which could be parametrically
the same size as the $2\chi/z$ term.  These corrections are related to
the $C_5/\omega z$ corrections we ignored in (\ref{eq:psi0thing1}),
and we sweep them under the rug here as well.

Comparison of (\ref{eq:psibdyWKB2}) with (\ref{eq:psibdyWKB}) identifies
\begin {equation}
  \eta_{\omega\rstar>0} =
  C_5
  \begin{pmatrix}
     -i e^{-i\m\pi/2} e^{i\omega\rstaro} \\
     e^{i\m\pi/2} e^{-i\omega\rstaro}
  \end {pmatrix}
  \equiv
  \begin{pmatrix}
     \eta^+_{\omega\rstar>0} \\
     \eta^-_{\omega\rstar>0}
  \end {pmatrix}
\label {eq:etabdy5}
\end {equation}
and then
\begin {equation}
  \chi_{\omega\rstar>0} =
  \frac{C_0}{2\omega}
  \begin{pmatrix}
     -i e^{-i\m\pi/2} e^{i\omega\rstaro} \\
     -e^{i\m\pi/2} e^{-i\omega\rstaro}
  \end{pmatrix} .
\end {equation}
(The net effect of the various corrections that we ignored above
can be absorbed into
a redefinition of the $C_0$ in the last formula---a redefinition
which  will not be
relevant since $C_0$ is so far arbitrary.)
Using the definition (\ref{eq:chihat}) of $\hat\chi$,
\begin {equation}
  \hat\chi_{\omega\rstar>0} =
  \hat C_0
  \begin{pmatrix}
     -i e^{-i\m\pi/2} e^{i\omega\rstaro} \\
     -e^{i\m\pi/2} e^{-i\omega\rstaro}
  \end{pmatrix}
  \equiv
  \begin{pmatrix}
     \hat\chi^+_{\omega\rstar>0} \\
     \hat\chi^-_{\omega\rstar>0}
  \end {pmatrix} ,
\label {eq:chihatbdy}
\end {equation}
where $\hat C_0 \equiv 2iC_0/(d-1)$.

% ----------------------------------------------------------------------------

\subsection{Putting it together}

The horizon condition for the quasi-normal modes, that there be no
$e^{+i\omega\rstar}$ components at the horizon, requires
$\eta^{(-)}$ and $\chi^{(-)}$ to vanish along the $\omega\rstar<0$ Stokes line.
From (\ref{eq:etaxi3}), this condition gives
\begin {equation}
   0
   =
   \begin{pmatrix} \hat\chi^{(-)} \\ \eta^{(-)} \end{pmatrix}_{\omega\rstar>0} +
   \left[
      \lambda \, \frac{k/\pi T}{(\omega/\pi T)^a}
      \tau_1 \utilde{\hat{\cal M}}
    + \Lambda \, \frac{\m}{(\omega/\pi T)^{1-a}}
      \tau_2 \utilde{\hat{\cal N}}
   \right]
   \begin{pmatrix} \hat\chi^{(+)} \\ \eta^{(+)} \end{pmatrix}_{\omega\rstar>0} .
\end {equation}
In terms of components,
\begin {equation}
   0
   =
   \begin{pmatrix} \hat\chi^- \\ \eta^- \end{pmatrix}_{\omega\rstar>0} +
   \left[
      \lambda \, \frac{k/\pi T}{(\omega/\pi T)^a}
      \utilde{\hat{\cal M}}
    + i \Lambda \, \frac{\m}{(\omega/\pi T)^{1-a}}
      \utilde{\hat{\cal N}}
   \right]
   \begin{pmatrix} \hat\chi^+ \\ \eta^+ \end{pmatrix}_{\omega\rstar>0} .
\end {equation}
Eqs.\ (\ref{eq:etabdy5}) and (\ref{eq:chihatbdy}) then give
\begin {equation}
  \left[
    i \lambda \, \frac{k/\pi T}{(\omega/\pi T)^a} \utilde{\hat{\cal M}}
    - \Lambda \, \frac{\m}{(\omega/\pi T)^{1-a}} \utilde{\hat{\cal N}}
  \right]
  e^{-i\m\pi} e^{i2\omega\rstaro}
  \begin {pmatrix}
    \hat C_0
    \\
    C_5
  \end {pmatrix}
  =
  \begin {pmatrix}
    - \hat C_0
    \\
    C_5
  \end {pmatrix} ,
\end {equation}
and so
\begin {equation}
  \left[
    i \lambda \, \frac{k/\pi T}{(\omega/\pi T)^a} \utilde{M}
    - \Lambda \, \frac{\m}{(\omega/\pi T)^{1-a}} \utilde{N}
  \right]
  e^{-i\m\pi} e^{i2\omega\rstaro}
  \begin {pmatrix}
    \hat C_0
    \\
    C_5
  \end {pmatrix}
  =
  \begin {pmatrix}
    \hat C_0
    \\
    C_5
  \end {pmatrix} ,
\end {equation}
where
\begin {align}
  \utilde{M} &\equiv
  \begin{pmatrix} -1 & 0 \\ 0 & 1 \end{pmatrix}
  \utilde{\hat{\cal M}}
  =
   -\frac{1}{d-1}
   \begin{pmatrix}
      1
      & d{-}2
      \\[4pt]
      d{-}2
      &
      1
  \end{pmatrix}
  ,
\\
  \utilde{N} &\equiv
  \begin{pmatrix} -1 & 0 \\ 0 & 1 \end{pmatrix}
  \utilde{\hat{\cal N}}
  =
   -\frac{1}{d-1}
   \begin{pmatrix}
      1
      & d
      \\[4pt]
      d
      &
      1
  \end{pmatrix}
  .
\end {align}
The quasinormal mode condition is then that
\begin {equation}
  \left[
    i \lambda \, \frac{k/\pi T}{(\omega/\pi T)^a} \utilde{M}
    - \Lambda \, \frac{\m}{(\omega/\pi T)^{1-a}} \utilde{N}
  \right]
  e^{-i\m\pi} e^{i2\omega\rstaro}
\end {equation}
have an eigenvalue equal to 1.
The eigenvectors are
\begin {equation}
  \begin{pmatrix} \hat C_0 \\ C_5 \end {pmatrix}
  =
  \begin{pmatrix} 1 \\ -1 \end {pmatrix}
  ~ \mbox{and} ~
  \begin{pmatrix} 1 \\ 1 \end {pmatrix} ,
\end {equation}
and the corresponding conditions are
\begin {equation}
  \left[
    \frac{(d-3)}{(d-1)} \, i \lambda \, \frac{k/\pi T}{(\omega/\pi T)^a}
    - \Lambda \, \frac{\m}{(\omega/\pi T)^{1-a}}
  \right]
  e^{-i\m\pi} e^{i2\omega\rstaro}
  = 1
\end {equation}
and
\begin {equation}
  \left[
    -i \lambda \, \frac{k/\pi T}{(\omega/\pi T)^a}
    + \frac{(d+1)}{(d-1)} \, \Lambda \, \frac{\m}{(\omega/\pi T)^{1-a}}
  \right]
  e^{-i\m\pi} e^{i2\omega\rstaro}
  = 1 ,
\end {equation}
respectively.
This is just the spin-$\tfrac12$ condition (\ref{eq:condition0})
with $\lambda$ and $\Lambda$ replaced by either
(i) $(d-3)\lambda/(d-1)$ and $\Lambda$ or
(ii) $-\lambda$ and $-(d+1)\Lambda/(d-1)$.
Making these substitutions in the Dirac result
(\ref{eq:dirac}) for the asymptotic quasinormal mode frequencies
yields the first and second branch spin-$\tfrac32$ results of
(\ref{eq:WKBbranch1}) and (\ref{eq:WKBbranch2}), respectively.

% ============================================================================

\section {Numerics}
\label {sec:numerics}

We would like to compare our analytic asymptotic formulas for
quasinormal mode frequencies to precise numerical results.
For numerics,
we will generalize the method used for the
spin-$\tfrac12$ case in ref.\ \cite{dirac}, which in turn was
inspired by methods used by others.  We make no claim as to whether
our method is the most efficient, but it gets the job done.%
\footnote{
  For an example of another method that has been applied to
  the spin-$\tfrac32$ problem,
  see the analysis of quasinormal
  modes for low-temperature Reissner-Nordstr\"om-AdS$_4$ black holes
  in ref.\ \cite{Gauntlett}.
}
Here we will focus only on the non-transverse modes, since the
transverse ones are described by the Dirac equation and so can be
treated numerically by the exact same procedure as ref.\ \cite{dirac}.
We will start out with a general discussion but will later
specialize to the case of $D{=}5$.

% -------------------------------------------------------------------------

\subsection{A 2nd-order equation}

Start from the basic equations of motion (\ref{eq:etaxi1}) in terms
of $\chi$ and $\eta$, which we write here in the form
\begin {equation}
   \partial_{\rstar} \vec V =
   f^{1/2} e^{2i\omega\rstar\tau_3}
   \left( k \tau_1 + \frac{\m}{z} \, \tau_2 \right)
   \utilde{\cal R} \vec V ,
\end {equation}
where
\begin {equation}
   \vec V \equiv \begin{pmatrix} \chi \\ \eta \end{pmatrix} ,
\qquad
   \utilde{\cal R} \equiv
   \begin {pmatrix}
      1-\frac{zg}{f} & -\frac{z}{f}\\[4pt]
      -g(2 - \frac{zg}{f}) & -(1 - \frac{zg}{f})
   \end {pmatrix}
   = \utilde{\cal R}^{-1} ,
\end {equation}
and
\begin {equation}
   g \equiv \int \frac{f'}{z} \,.
\end {equation}
In terms of the $\tau_3 = \pm 1$ components $\chi^\pm$ and
$\eta^\pm$ of the spinors
$\chi$ and $\eta$, this is
\begin {align}
   \partial_{\rstar} \vec V^+ &=
   f^{1/2} e^{2i\omega\rstar}
   \left( k - i \, \frac{\m}{z} \right)
   \utilde{\cal R} \vec V^- ,
\label {eq:Vp}
\\
   \partial_{\rstar} \vec V^- &=
   f^{1/2} e^{-2i\omega\rstar}
   \left( k + i \, \frac{\m}{z} \right)
   \utilde{\cal R} \vec V^+ ,
\label {eq:Vm}
\end {align}
where
\begin {equation}
   \vec V^\pm \equiv \begin{pmatrix} \chi^\pm \\ \eta^\pm \end{pmatrix}
   .
\end {equation}
Solving (\ref{eq:Vp}) for $\vec V^-$ and then plugging into
(\ref{eq:Vm}) gives
\begin {equation}
   \partial_{\rstar} \left[
     f^{-1/2} e^{-2i\omega\rstar}
     \left( k - i \, \frac{\m}{z} \right)^{-1}
     \utilde{\cal R}^{-1}
     \partial_{\rstar} \vec V^+
   \right]
   =
   f^{1/2} e^{-2i\omega\rstar}
   \left( k + i \, \frac{\m}{z} \right)
   \utilde{\cal R} \vec V^+ ,
\end {equation}
which may be rewritten as
\begin {equation}
   f \partial_z^2 \vec V^+
   + \left(
        2i \omega + \utilde{\cal S}
        + \frac{f'}{2}
        - \frac{i\m f}{z(kz-i\m)}
     \right)
     \partial_z \vec V^+
   - \left(
        \frac{\m^2}{z^2} + k^2
     \right) \vec V^+
   = 0 ,
\label {eq:calnice}
\end {equation}
where
\begin {equation}
  \utilde{\cal S} \equiv
  f \utilde{\cal R}^{-1} \partial_z ({\cal R}^{-1}) =
  \begin {pmatrix}
     f'-g & -1 \\
     g^2 + \frac{2 f f'}{z} - 2 g f' & -(f'-g)
  \end {pmatrix} .
\end {equation}
We will find it useful to rewrite this as
\begin {equation}
  \utilde{\cal S} = (f'-g) \utilde{\Sigma}_3
    - \utilde{\Sigma}_+
    + \left[ g^2 + \frac{2 f f'}{z} - 2 g f' \right] \utilde{\Sigma}_- ,
\end {equation}where the $\utilde{\Sigma}_i$ are Pauli matrices and
$\utilde{\Sigma}_\pm \equiv
 \tfrac12 (\utilde{\Sigma}_1 \pm i \utilde{\Sigma}_2)$.
For $D=5$,
\begin {equation}
  \utilde{\cal S}
  = -\tfrac83 \, z^3 \utilde{\Sigma}_3
    - \utilde{\Sigma}_+
    - 8 z^2 ( 1 + \tfrac19 \, z^4 ) \utilde{\Sigma}_- .
\end {equation}
We now factor out the behavior at the boundary that we do not want
for quasinormal modes by rescaling
\begin {equation}
   \vec V^+ = z^{-\m} \vec H .
\label {eq:Hdef}
\end {equation}
That is, the condition for quasinormal modes is $\vec H(0) = 0$.
Substituting (\ref{eq:Hdef}) into (\ref{eq:calnice}),
the equation for $\vec H$ is%
\footnote{
   For $\utilde{\cal S}=0$, (\ref{eq:Hnice}) would be identical to
   the spin-$\tfrac12$ equation (3.9) of ref.\ \cite{dirac}.
   Our (\ref{eq:Hnice}) is simply that equation with
   $i\omega$ replaced by $i\omega + \tfrac12 \utilde{\cal S}$.
}
\begin {multline}
   z f \partial_z^2 \vec H
   + \left(
        (2i \omega + \utilde{\cal S})z
        - 2 \m f
        + \frac{zf'}{2}
        - \frac{i\m f}{kz-i\m}
     \right)
     \partial_z \vec H
\\
   - \left[
        \frac{\m^2(1-f)}{z}
        + \m \left(
              (2i\omega + \utilde{\cal S})
              + \frac{f'}{2}
              - \frac{k f}{kz-i\m}
            \right)
        + k^2 z
     \right] \vec H
   = 0 .
\label {eq:Hnice}
\end {multline}

% -------------------------------------------------------------------------

\subsection{A recursion relation}

To implement the correct infalling boundary condition at the horizon, we
require $\vec V^+$ to be regular at the horizon, which will ensure via
(\ref{eq:WKB}) that the corresponding components of $\psi_0$ and $\psi_5$
behave like $e^{-i\omega\rstar}$ at the horizon and not $e^{+i\omega\rstar}$.%
\footnote{
   The $\tau_3{=}-1$ components then follow suit: If $\vec V^+$ is
   regular at the horizon, then (\ref{eq:Vm}) implies that
   $\vec V^-$ will behave like $e^{-2i\omega\rstar}$ there, and
   (\ref{eq:WKB}) then implies that $\psi_0^-$ and $\psi_5^-$ will
   behave like $e^{-i\omega\rstar}$, as desired.
}
So, we take $\vec V^+$ to have a power series solution around
$z=\zh$.  Working in units where $\zh = 1$,
\begin {equation}
   \vec H(z) = \sum_{n=0}^\infty \vec a_n (1-z)^n .
\label {eq:Hx}
\end {equation}
By plugging this series into the equation of motion (\ref{eq:Hnice}),
we find a linear recursion relation for the coefficients $\vec a_n$
of the form
\begin {equation}
   \sum_{j=0}^{j_{\rm max}} \utilde\alpha_{-j}(n) \, \vec a_{n-j} = 0 ,
\label {eq:recurse0}
\end {equation}
with the understanding that $\vec a_n$ vanishes for negative $n$.
For $D{=}5$, (\ref{eq:recurse0})
is a 9-term recursion relation ($j_{\rm max}{=}8$)
with explicit coefficients given in appendix \ref{app:coeffs}.
For $D{=}4$, it is a 7-term recursion relation, also given
in the appendix.
The $\utilde\alpha_{-j}(n)$ above depend on $\omega$, $\k$,
and $\m$.

Solutions exist for arbitrary values of $\vec a_0$, to which the
other coefficients are then linearly related by
(\ref{eq:recurse0}).  Define $\utilde A_n$ by writing
\begin {equation}
  \vec a_n = \utilde A_n \vec a_0 .
\end {equation}
Then $\utilde A_n$ satisfies the same recursion relation
(\ref{eq:recurse0}) that $\vec a_n$ does,
\begin {subequations}
\label {eq:recurse1}
\begin {equation}
   \sum_{j=0}^{j_{\rm max}} \utilde\alpha_{-j}(n) \, \utilde A_{n-j} = 0 ,
\end {equation}
initialized by
\begin {equation}
   \utilde A_0 = \openone .
\end {equation}
\end {subequations}
The series solution (\ref{eq:Hx}) may then be written as
\begin {equation}
   H(z) = \utilde{\cal H}(z) \, \vec a_0
\end {equation}
with
\begin {equation}
   \utilde {\cal H}(z) \equiv
   \sum_{n=0}^\infty \utilde A_n (1-z)^n .
\label {eq:calH}
\end {equation}

A quasinormal mode solution corresponds to $H(1)=0$ for some value
of $\vec a_0$.  The existence of such an $\vec a_0$
is equivalent to the condition that
\begin {equation}
   \det \utilde{\cal H}(1) = 0 .
\label {eq:detH}
\end {equation}
Our strategy for numerics is to use the recursion relation
(\ref{eq:recurse1}) and series (\ref{eq:calH}), cut off at some
suitably high order $n_{\rm max}$, to compute
$\det \utilde{\cal H}(1)$ for a given choice of $\omega$
(and fixed $k$ and $\m$).  We then scan the complex $\omega$ plane
to find the quasinormal frequencies, corresponding to the zeros
of $\det \utilde{\cal H}(1)$.  All calculations are carried out with
very high precision arithmetic, with the precision and $n_{\rm max}$
increased as necessary to achieve numerical stability and the
desired numerical accuracy.

% -------------------------------------------------------------------------

\subsection{Numerical test of asymptotic formulas}

\begin {figure}
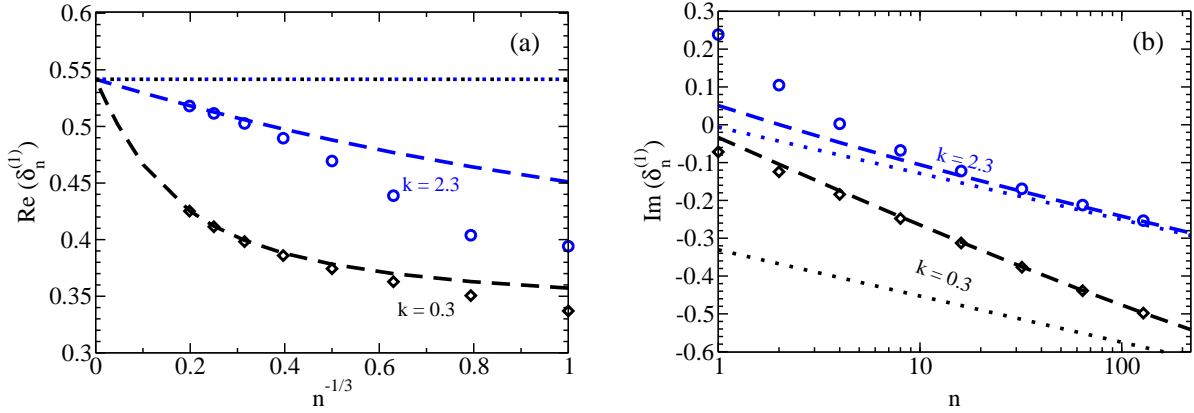

\begin {center}
  \includegraphics[scale=0.30]{tabre1.eps}
  \hspace{0.2in}
  \includegraphics[scale=0.30]{tabim1.eps}
  \caption{
     \label{fig:deltan1}
     A plot of the (a) real and (b) imaginary parts of
     the first-branch offsets $\delta_n^{(1)}$ defined by (\ref{eq:deltan}) for
     $D=5$, $m\Rads=\tfrac32$, $k/\pi T = 0.3$ and $2.3$,
     and the representative overtones $n=1,2,4,8,16,\cdots,128$.
     Note the different choices made for the horizontal axis in the
     two figures, and that large $n$ corresponds to the left
     and right hand sides of (a) and (b) respectively.
     Data points are from our numerics, whereas the dashed lines
     indicate the value of $\delta_n$ that would be
     given by the asymptotic formula (\ref{eq:WKBbranch1}).
     Dotted lines are the same formula but dropping the $\m$ terms in
     the argument of the logarithms.
  }
\end {center}
\end {figure}

We have already shown a comparison of numerical results to
asymptotic formulas for quasinormal mode frequencies in
fig.\ \ref{fig:wplane}.  In order to perform a more precise comparison,
it is
useful to focus on the offset $\delta$ from the leading
$O(n)$ asymptotic formula, defined by
\begin {equation}
  \delta_n \equiv \frac{\omega_n}{\Delta\omega_\infty} - n
  = \frac{\omega_n}{(2-2i)\pi T} - n
\label {eq:deltan}
\end {equation}
for $D{=}5$, where $\Delta\omega_\infty$ is the $n{\to}\infty$
spacing between consecutive modes on a given branch.
Here we will focus on quasinormal mode frequencies in the
right-half complex $\omega$ plane.
For the first branch of non-transverse quasinormal mode solutions,
fig.\ \ref{fig:deltan1} shows data points for $\delta_{n}$
from numerics, plotted against dashed lines showing the
asymptotic results taken from (\ref{eq:WKBbranch1}).
The asymptotic formula works very well at large $n$.

The dotted lines in fig.\ \ref{fig:deltan1}
indicate what happens if one were to drop the
$\m$ terms in the argument of the logarithm
in the asymptotic formula (\ref{eq:WKBbranch1}).
We see that including these terms has significantly improved the
accuracy for small $k$ and moderate $n$.

% =========================================================================
% =========================================================================

\begin{acknowledgments}

This work was supported, in part, by the U.S. Department
of Energy under Grant No.~DE-SC0007984.

\end{acknowledgments}

% =========================================================================
% =========================================================================

\appendix

% =========================================================================

\section{Gravitino and ghost determinants in $(d{+}1)$-dimensional
         Einstein spaces}
\label {app:ghost}

Since one of the possible applications of this work is to compute $1/\Nc^2$
corrections to the free energy of strongly-coupled large-$\Nc$
${\cal N}{=}4$ super Yang Mills gauge
theory at finite temperature via supergravity loops, in this appendix
we show how to account for the gravitino gauge degrees of freedom.
More precisely we construct the gravitino and ghost determinants.

To maintain full generality we work in a $(d{+}1)$-dimensional asymptotically
AdS space, with the AdS radius denoted by $L$.
We begin by decomposing the gravitino field into fields which are
divergence-free
and gamma-traceless ($\Psi^\perp_M$), fields which are a pure trace
($\phi$), and fields which reflect the gravitino gauge symmetry%
\footnote{
  One can check that $\delta \Psi_M=D_M\epsilon - \frac{1}{2L}\Gamma_M\epsilon$
  is a symmetry of the equation of motion
  $\Gamma^{MNP} D_N\Psi_P + \frac{d-1}{2L} \Gamma^{MN}\Psi_N=0$.
}
 ($\zeta$):
\begin{equation}
  \Psi_M =
  \Psi_M^\perp
  + \frac{1}{d+1}\Gamma_M \phi
  + \Bigl(D_M-\frac{\Gamma_M}{2L}\Bigr) \zeta,
\end{equation}
where $\Psi_M^\perp$ obeys
\begin{equation}
  D^M\Psi_M^\perp=0, \;\;\;\;\Gamma^M\Psi_M^\perp=0.
\end{equation}
(Note that the superscript $\perp$ denotes all divergence-free, gamma-traceless
$\Psi_M$'s and not just the subset referred to as ``transverse'' in the main
text.)
The path integral measure over the spin $3/2$ and $1/2$ fields is defined as
\begin{equation}
  \int{\cal D} \Psi_M \>
  \exp\bigg(-\int d^{d+1}x \> \sqrt{-g} \, \bar \Psi^M\Psi_M\bigg)=1
\end{equation}
and
\begin{equation}
  \int{\cal D}\phi \>
  \exp\bigg(-\int d^{d+1} x \> \sqrt{-g} \, \bar\phi \phi\bigg)=1.
\end{equation}
The change of integration from $\Psi_M$ to $\Psi_M^\perp$, $\phi$ and $\zeta$ is
accompanied by a Jacobian, $Z_{\text{ghost}}$, which can be found from
\begin{eqnarray}
  1 &=& \int {\cal D} \Psi_M \>
        \exp\bigg(-\int d^{d+1}x \> \sqrt{-g} \, \bar \Psi^M\Psi_M\bigg)
\nonumber\\
    &=& \int {\cal D}\Psi_M^\perp \int {\cal D}\phi
        \int {\cal D}\zeta \> Z_{\text{ghost}}
        \exp\bigg[
           -\bar \Psi^{\perp M}\Psi_M^\perp
           +\frac{1}{d+1}\bar\phi\phi
        \bigg]
\nonumber\\
   && \times
      \exp\bigg[
          \frac{1}{d+1}\bar\phi
            \Gamma^M\Bigl(D_M-\frac{1}{2L}\Gamma_M\Bigr)\zeta
         -\frac{1}{d+1} \,
             \Bigl(D_M\bar\zeta+\frac{1}{2L} \bar\zeta \Gamma_M\Bigr)
             \Gamma^M\phi
      \bigg]
\nonumber\\
   && \times
      \exp\bigg[
         -\Bigl(D_M\bar\zeta+\frac{1}{2L}\bar \zeta \Gamma_M\Bigr)
          \Bigl(D^M-\frac{1}{2L}\Gamma^M\Bigr)\zeta
      \bigg],
\label{jacobian1}
\end{eqnarray}
where we introduced the notation
\begin{equation}
  D_M\bar\zeta=\partial_M\bar\zeta -\frac14 \bar\zeta \omega_M^{mn}\gamma^{mn}.
\end{equation}
After a further redefinition with unit Jacobian
\begin{equation}
  \phi'=\phi+\Gamma^M\Bigl(D_M-\frac{1}{2L}\Gamma_M\Bigr)\zeta
\end{equation}
and after performing some of the integrals in (\ref{jacobian1}) we are left with
\begin{equation}
  1 =
  \int{\cal D}\zeta \> Z_{\text{ghost}}
  \exp\bigg[-
     \Bigl(D_M\bar \zeta-\frac 1{d+1} D_N\bar\zeta \Gamma^N\Gamma_M\Bigr)
     \Bigl(D^M\zeta-\frac 1{d+1} \Gamma^M\Gamma^P D_P\zeta\Bigr)
  \bigg].
\label{jacobian2}
\end{equation}
After integration by parts, some Dirac algebra manipulations
together with $[D_M,D_N]\zeta=\tfrac 14 R_{MNpq}\gamma^{pq}\zeta$ and
$R_{MN}=-\tfrac d{L^2} g_{MN}$, (\ref{jacobian2}) yields
\begin{equation}
  1 =
  \int{\cal D}\zeta \> Z_{\text{ghost}}
  \exp\bigg[
    \frac d{d+1}
    \bigg(\bar\zeta D^M D_M \zeta- \frac{d+1}{4L^2}\bar\zeta\zeta\bigg)
  \bigg].
\end{equation}
The Jacobian (ghost determinant) is now determined to be
\begin{equation}
  Z_{\text{ghost}}=\frac1{\text{det}_{\tfrac12}(D^M D_M - \frac{d+1}{4L^2})},
\end{equation}
where the subscript $\tfrac 12$ in the above expression is meant to express
that the operator $D^M D_M - \frac{d+1}{4}$ acts on spin-$\tfrac 12$ fields.

%% Note to us: strictly speaking I get the determinant of the differential
%% operator above times a numerical factor.

On the other hand, integrating out the gravitino from its quadratic action
yields
\begin{eqnarray}
  Z_{\text{gravitino}}
  &\propto& \int {\cal D}\Psi_M \>
  \exp\bigg[
    \int d^{d+1} x \> \sqrt{-g} \, \bar\Psi_M
    \Bigl(\Gamma^{MNP}D_N\Psi_P+\frac{d-1}{2L}\Gamma^{MN}\Psi_N\Bigr)
  \bigg]
\nonumber\\
  &=&
  \int{\cal D}\Psi^{\perp}_M \int {\cal D} \phi \int{\cal D}\zeta \> Z_{\text{ghost}}
  \exp\bigg[
    \int d^{d+1} x \> \sqrt{-g} \, \bar\Psi_M^\perp
    \Bigr(\Gamma^{N}D_N-\frac{d-1}{2L}\Bigr)\Psi_M^\perp
  \bigg]
\nonumber\\
  &&\qquad\times
  \exp\bigg[
    -\frac{d(d-1)}{(d+1)^2}\int d^{d+1} x \> \sqrt{-g} \, \bar\phi
    \Bigl(\Gamma^M D_M+\frac{d+1}{2L}\Bigr)\phi
  \bigg]
\nonumber\\
\end{eqnarray}
Throwing away the volume of the gauge group yields
\begin{equation}
  Z_{\text{gravitino}} =
  \frac{\text{det}_{\tfrac32,\perp}(\Gamma^M D_M-\frac{d-1}{2L} )
           \;\text{det}_{\tfrac 12} (\Gamma^M D_M +\frac{d+1}{2L})}
       {\text{det}_{\tfrac12}(D^M D_M - \frac{d+1}{4L^2})},
\label{gravitino_part}
\end{equation}
where the $\text{det}_{\tfrac32 , \perp}$ denotes the evaluation of the
determinant on the subspace of spin-$\tfrac32$ fields which are
divergence-free
and gamma-traceless.

The gravitino partition function can be rewritten in a more compact form
by using the same sort of manipulations as before to see that
\begin {equation}
   \text{det}_{\tfrac12}(D^M D_M - \tfrac{d+1}{4L^2})
   =
   \text{det}_{\tfrac12}(\Gamma^M D_M + \tfrac{d+1}{2L}) \;
   \text{det}_{\tfrac12}(\Gamma^M D_M - \tfrac{d+1}{2L})
\end {equation}
and so%
\footnote{
  The results presented in this appendix are rather
  straightforward generalizations of Zhang and Zhang \cite{ZhangZhang}
  to dimensions
  other than $D{=}3$. However, this is as far as we can go in
  simplifying the gravitino partition function in a generic Einstein space.
  That is because, in order to cast the spin 3/2 determinant as the
  square root of a Laplacian as in \cite{ZhangZhang}, one needs to use the
  Riemann curvature tensor. Only in maximally symmetric spaces such as
  AdS is the latter simply expressed in terms of the metric as
  $R_{IJKL} \propto g_{IK} g_{JL} - g_{IL} g_{JK}$, leading to the
  further simplification obtained by Zhang and Zhang.
  (For another discussion of the gravitino determinant in the specific
  case of pure AdS, see also ref.\ \cite{AdSdavid}.)
}
%%\begin{equation}
%%  Z_{\text{gravitino}} =
%%  \frac{\text{det}_{\tfrac32,\perp}(\Gamma^M D_M-\frac{d-1}{2L} )}
%%       {\sqrt{\text{det}_{\tfrac12}(D^M D_M - \frac{d+1}{4L^2})}},
%%\end{equation}
\begin{equation}
  Z_{\text{gravitino}} =
  \frac{\text{det}_{\tfrac32,\perp}(\Gamma^M D_M-\frac{d-1}{2L} )}
       {\text{det}_{\tfrac12}(\Gamma^M D_M - \frac{d+1}{2L} )} \,.
\label {eq:Zratio}
\end{equation}
Now recall from the main text that one of the branches of the
gravitino quasinormal modes had the same frequencies as the
ghost quasinormal modes.
The form (\ref{eq:Zratio}) shows that, if one writes each determinant following
Denef-Hartnoll-Sachdev as a product of factors involving quasinormal
frequencies \cite{DHS}, then the contributions from this branch
will exactly cancel the ghost determinant.

% =========================================================================

\section{Recursion coefficients \boldmath${\alpha}_{-j}(n)$}
\label {app:coeffs}

\subsection{\boldmath$D=5$}

For $D{=}5$, the results for the coefficients of the recursion relation
(\ref{eq:recurse0}) are
\begin {equation}
   \utilde\alpha_{-j}(n)
   = \alpha_{-j}^0(n) \, \utilde\openone
   + \alpha_{-j}^3(n) \, \utilde\Sigma_3
   + \alpha_{-j}^+(n) \, \utilde\Sigma_+
   + \alpha_{-j}^-(n) \, \utilde\Sigma_- ,
\end {equation}
where $\alpha_{-j}^0$ are the same as the coefficients for the
spin-$\tfrac12$ case \cite{dirac},
\begin {subequations}
\begin {align}
   \alpha_{-6}^0 &= \alpha_{-7}^0 = \alpha_{-8}^0 = 0 ,
\\
   \alpha_{-5}^0 &=
   -(4+\m-n)(5+\m-n)k ,
\\
   \alpha_{-4}^0 &=
   (4+\m-n)\bigl[ (20+4\m-6n)k + (-2-\m+n)i\m \bigr] ,
\\
   \alpha_{-3}^0 &=
   (-120-54\m-6\m^2+85n+20\m n-15n^2)k - k^3
\nonumber\\ & \qquad
   + (24+18\m+3\m^2-23n-8\m n+5n^2)i\m ,
\\
   \alpha_{-2}^0 &=
   (80+36\m+4\m^2-80n-20\m n+20n^2)k + 2k^3
\nonumber\\ & \qquad
   + (-24-18\m-3\m^2+32n+12\m n-10n^2)i\m
   - i k^2\m + (4+2\m-2n)i k \omega ,
\\
   \alpha_{-1}^0 &=
   (-18-6\m-\m^2+32n+8\m n-14n^2)k - k^3
\nonumber\\ & \qquad
   + (8+6\m+\m^2-18n-8\m n+10n^2)i\m
\nonumber\\ & \qquad
   + i k^2\m + (-2-2\m+2n)\m \omega + (-4-2\m+4n)i k \omega ,
\\
   \alpha_{0}^0 &=
   2n(k-i\m)(-1+2n-i \omega) ,
\end {align}
\end {subequations}
and the others are
\begin {subequations}
\begin {align}
   \alpha_{-6}^3 &= \alpha_{-7}^3 = \alpha_{-8}^3 = 0 ,
\\
   \alpha_{-5}^3 &=
   \tfrac83 (5+\m-n)k ,
\\
   \alpha_{-4}^3 &=
   -\tfrac83(20+4\m-5n)k +\tfrac83(4+\m-n)i\m ,
\\
   \alpha_{-3}^3 &=
   \tfrac{16}3(15+3\m-5n)k - \tfrac83(12+3\m-4n)i\m ,
\\
   \alpha_{-2}^3 &=
   -\tfrac{16}3(10+2\m-5n)k + 8(4+\m-2n)i\m ,
\\
   \alpha_{-1}^3 &=
   \tfrac83(5+\m-5n)k - \tfrac83(4+\m-4n)i\m ,
\\
   \alpha_{0}^3 &=
   \tfrac83\,n(k-i\m) ,
\end {align}
\end {subequations}
and
\begin {subequations}
\begin {align}
   \alpha_{-3}^+ &= \alpha_{-4}^+ = \alpha_{-5}^+ = \alpha_{-6}^0
   = \alpha_{-7}^0 = \alpha_{-8}^0 = 0 ,
\\
   \alpha_{-2}^+ &=
   -(2+\m-n)k ,
\\
   \alpha_{-1}^+ &=
   (2+\m-2n)k - (1+\m-n)i\m ,
\\
   \alpha_{0}^+ &=
   n(k-i\m) ,
\end {align}
\end {subequations}
and
\begin {subequations}
\begin {align}
   \alpha_{-8}^- &=
   -\tfrac89(8+\m-n)k ,
\\
   \alpha_{-7}^- &=
   \tfrac89(56+7\m-8n)k - \tfrac89(7+\m-n)i\m ,
\\
   \alpha_{-6}^- &=
   -\tfrac{56}9(24+3\m-4n)k + \tfrac89(42+6\m-7n)i\m,
\\
   \alpha_{-5}^- &=
   \tfrac{56}9(40+5\m-8n)k - \tfrac83(35+5\m-7n)i\m ,
\\
   \alpha_{-4}^- &=
   -\tfrac89(316+44\m-79n)k + \tfrac{40}9(28+4\m-7n)i\m ,
\\
   \alpha_{-3}^- &=
   \tfrac{32}9(69+12\m-23n)k - \tfrac{32}9(33+6\m-11n)i\m ,
\\
   \alpha_{-2}^- &=
   -\tfrac{16}9(82+17\m-41n)k + \tfrac{64}3(4+\m-2n)i\m ,
\\
   \alpha_{-1}^- &=
   \tfrac{16}9(22+5\m-22n)k - \tfrac{16}9(17+5\m-17n)i\m ,
\\
   \alpha_{0}^- &=
   \tfrac{80}9 \, n(k-i\m) .
\end {align}
\end {subequations}
Note that this is a 9-term recursion relation, whereas the spin-$\tfrac12$
case \cite{dirac} gave a
6-term recursion relation.  We are unsure if some
clever change of variables could yield simpler recursion
relations.

% -------------------------------------------------------------------------

\subsection{\boldmath$D=4$}

For the sake of completeness, we also give here the recursion coefficients
for the case of $D{=}4$.
The $\alpha_{-j}^0$ are again the same as the coefficients for the
spin-$\tfrac12$ case, which are
\begin{subequations}
\begin{align}
   \alpha_{-6}^0 &= \alpha_{-5}^0 = 0 ,
\\
   \alpha_{-4}^0 &=
   \tfrac12(7+2\m-2n)(4+\m-n)k ,
\\
   \alpha_{-3}^0 &=
  \tfrac12(-84-45\m-6\m^2+58n+16\m n-10n^2)k - k^3
\nonumber\\ & \qquad
   + \tfrac12(3+2\m-2n)(3+\m-n)i\m ,
\\
   \alpha_{-2}^0 &=
   \tfrac12(84+45\m+6\m^2-82n-24\m n+20n^2)k + 2k^3
\nonumber\\ & \qquad
   - \tfrac12(18+18\m+4\m^2-25n-12\m n+8n^2)i\m
   - i k^2\m + (4+2\m-2n)i k \omega ,
\\
   \alpha_{-1}^0 &=
   \tfrac12(-24-9\m-2\m^2+42n+12\m n-18n^2)k - k^3
\nonumber\\ & \qquad
   + \tfrac12(9+9\m+2\m^2-21n-12\m n+12n^2)i\m
\nonumber\\ & \qquad
   + i k^2\m + (-2-2\m+2n)\m \omega + (-4-2\m+4n)i k \omega ,
\\
   \alpha_{0}^0 &=
   \tfrac12n(k-i\m)(-3+6n-4i \omega) .
\end{align}
\end{subequations}
The other coefficients are
\begin{subequations}
\begin{align}
   \alpha_{-5}^3 &= \alpha_{-6}^3 = 0 ,
\\
   \alpha_{-4}^3 &=
   -\tfrac32 (4+\m-n)k ,
\\
   \alpha_{-3}^3 &=
   \tfrac32(12+3\m-4n)k -\tfrac32(3+\m-n)i\m ,
\\
   \alpha_{-2}^3 &=
   -\tfrac92(4+\m-2n)k + \tfrac32(6+2\m-3n)i\m ,
\\
   \alpha_{-1}^3 &=
   \tfrac32(4+\m-4n)k - \tfrac32(3+\m-3n)i\m ,
\\
   \alpha_{0}^3 &=
   \tfrac32\,n(k-i\m) ,
\end{align}
\end{subequations}
and
\begin{subequations}
\begin{align}
   \alpha_{-3}^+ &= \alpha_{-4}^+ = \alpha_{-5}^+ = \alpha_{-6}^0 = 0 ,
\\
   \alpha_{-2}^+ &=
   -(2+\m-n)k ,
\\
   \alpha_{-1}^+ &=
   (2+\m-2n)k - (1+\m-n)i\m ,
\\
   \alpha_{0}^+ &=
   n(k-i\m) ,
\end{align}
\end{subequations}
and
\begin{subequations}
\begin{align}
   \alpha_{-6}^- &=
   -\tfrac34(6+\m-n)k ,
\\
   \alpha_{-5}^- &=
   \tfrac34(30+5\m-6n)k - \tfrac34(5+\m-n)i\m ,
\\
   \alpha_{-4}^- &=
   -\tfrac{15}4(12+2\m-3n)k + \tfrac34(20+4\m-5n)i\m,
\\
   \alpha_{-3}^- &=
   \tfrac32(42+9\m-14n)k - \tfrac32(15+3\m-5n)i\m ,
\\
   \alpha_{-2}^- &=
   -\tfrac94(26+7\m-13n)k + \tfrac92(6+2\m-3n)i\m ,
\\
   \alpha_{-1}^- &=
   \tfrac94(10+3\m-10n)k - \tfrac94(7+3\m-7n)i\m ,
\\
   \alpha_{0}^- &=
   \tfrac{27}4 \, n(k-i\m) .
\end{align}
\end{subequations}
This is a 7-term recursion relation.

% =========================================================================
% =========================================================================

\end{document}